\documentclass[12pt,fleqn]{article}

\usepackage{amsmath,amsfonts,amssymb,cite}
\usepackage{graphicx}

\def\noi{\noindent}

\newcommand{\Title}[1]{\noi {{\Large\bf #1}}\\[1ex]}
\newcommand{\Author}[2]{\noi{\bf #1}\\[2ex]\noi{\normalsize\it #2}\\}
\newcommand{\foom}[1]{\protect\footnotemark[#1]}

\def\email#1#2{\footnotetext[#1]{e-mail: #2}\addtocounter{footnote}{1}}
\newcommand{\Abstract}[1]{\vskip 2mm \begin{center}
        \parbox{16.4cm}{\small\noi #1} \end{center}\medskip}

\def\noi{\noindent}

\def\nqq{\hspace*{-2em}}
\def\nhq{\hspace*{-0.5em}}

\def\qq{\hspace*{2em}}
\def\cm{\hspace*{1cm}}
\def\inch{\hspace*{1in}}

\def\Jl#1#2{#1 {\bf #2},\ }
\def\ApJ#1 {\Jl{Astroph. J.}{#1}}
\def\CQG#1 {\Jl{Class. Quantum Grav.}{#1}}
\def\DAN#1 {\Jl{Dokl. AN SSSR}{#1}}
\def\GC#1 {\Jl{Grav. Cosmol.}{#1}}
\def\GRG#1 {\Jl{Gen. Rel. Grav.}{#1}}
\def\JETF#1 {\Jl{Zh. Eksp. Teor. Fiz.}{#1}}
\def\JETP#1 {\Jl{Sov. Phys. JETP}{#1}}
\def\JHEP#1 {\Jl{JHEP}{#1}}
\def\JMP#1 {\Jl{J. Math. Phys.}{#1}}
\def\NPB#1 {\Jl{Nucl. Phys. B}{#1}}
\def\NP#1 {\Jl{Nucl. Phys.}{#1}}
\def\PLA#1 {\Jl{Phys. Lett. A}{#1}}
\def\PLB#1 {\Jl{Phys. Lett. B}{#1}}
\def\PRD#1 {\Jl{Phys. Rev. D}{#1}}
\def\PRL#1 {\Jl{Phys. Rev. Lett.}{#1}}

\def\al{&\nhq}
\def\lal{&&\nqq {}}
\def\eq{Eq.\,}
\def\eqs{Eqs.\,}
\def\beq{\begin{equation}}
\def\eeq{\end{equation}}
\def\bear{\begin{eqnarray}}
\def\bearr{\begin{eqnarray} \lal}
\def\ear{\end{eqnarray}}
\def\earn{\nonumber \end{eqnarray}}
\def\besub{\begin{subequations}}
\def\esub{\end{subequations}}

\def\nnn{\nonumber\\ \lal }

\def\yy{\\[5pt] {}}
\def\yyy{\\[5pt] \lal }
\def\eql{\al =\al}

\def\tst{\textstyle}

\def\fract#1#2{{\tst\frac{#1}{#2}}}

\def\half{{\fract{1}{2}}}

\def\e{{\,\rm e}}
\def\d{\partial}

\def\im{\mathop{\rm Im}\nolimits}

\def\sign{\mathop{\rm sign}\nolimits}
\def\diag{\mathop{\rm diag}\nolimits}

\def\const{{\rm const}}

\def\ep{\epsilon}

\def\then{\ \Rightarrow\ }

\newcommand{\vars}[1]{\left\{\begin{array}{ll}#1\end{array}\right.}

\usepackage{color}

\def\rf{\eqref}
\def\eqn#1{Eq.\,\rf{#1}}

\def\mn{_{\mu\nu}}
\def\MN{^{\mu\nu}}
\def\mN{_\mu^\nu}

\def\N{{\mathbb N}}
\def\R{{\mathbb R}}

\def\Z{{\mathbb Z}}
\def\cF{{\cal F}}

\def\da{\delta\alpha}
\def\db{\delta\beta}
\def\df{\delta\phi}
\def\dg{\delta\gamma}
\def\Veff{V_{\rm eff}}

\def\wh{wormhole}
\def\whs{wormholes}
\def\bh{black hole}
\def\bhs{black holes}

\def\sph{spherically symmetric}
\def\ssph{static, spherically symmetric}
\def\asflat{asymptotically flat}
\def\asdS{asymptotically de Sitter}
\def\asAdS{asymptotically anti-de Sitter}
 
\def\KS{Kan\-tow\-ski-Sachs}
\def\Schr{Schr\"o\-din\-ger}
\def\Scz{Schwarz\-schild}

\begin{document}
\thispagestyle{empty}

\Title{Scalar fields as sources for wormholes and regular black holes}

\Author{Kirill A. Bronnikov\foom 1}
 	{\small
	 VNIIMS, Ozyornaya ul. 46, Moscow 119361, Russia;\\
	 Institute of Gravitation and Cosmology, Peoples' Friendship University of Russia (RUDN University),\\ 
           \qq  ul. Miklukho-Maklaya 6, Moscow 117198, Russia;\\
	National Research Nuclear University ``MEPhI'', Kashirskoe sh. 31, Moscow 115409, Russia}

\Abstract
  {We review nonsingular \ssph\ solutions of general relativity with minimally coupled scalar
   fields. Considered are wormholes and regular black holes (BHs) without a center, including black universes 
   (BHs with expanding cosmology beyond the horizon). Such configurations require a ``ghost'' field 
   with negative kinetic energy $K$. Ghosts can be invisible under usual conditions if $K<0$ only in strong-field 
   region (``trapped ghost''), or they rapidly decay at large radii. Before discussing particular examples, 
   some general results are presented, such as the necessity of anisotropic matter for asymptotically flat 
   or AdS \whs, no-hair and global structure theorems for BHs with scalar fields. The stability properties of 
   scalar \whs\ and regular BHs under spherical perturbations are discussed. It is stressed that the effective
   potential $\Veff$ for perturbations has universal shapes near generic \wh\ throats (a positive pole 
   regularizable by a Darboux transformation) and near transition surfaces from canonical to ghost 
   scalar field behavior (a negative pole at which the perturbation finiteness requirement plays a stabilizing role).
   Positive poles of $\Veff$ emerging at ``long throats'' (with the radius 
   $r \approx r_0 + \const \cdot x^{2n}$, $n > 1$, $x=0$ is the throat) may be regularized by repeated
   Darboux transformations for some values of $n$. 
   }

] 
\email 1 {kb20@yandex.ru}

\section{Introduction}

  Space-time singularities exist in a great number of solutions of general relativity (GR) with or without various
  material sources, and at each of them the theory itself demonstrates conditions under which it cannot work 
  any more. However, construction of various nonsingular solutions of GR, which are especially interesting 
  and attractive, is also a long-term tradition. In particular, nonsingular \ssph\ space-times, which are a subject 
  of this paper, may be classified as follows: (i) starlike (or solitonic, or particlelike) space-times with a regular
  center, (ii) black holes (BHs) with a regular center, (iii) space-times having no center and no horizons, including
  wormholes and some other geometries like horns and flux tubes, and (iv) space-times without a center 
  but containing horizons, and among them are a few classes of regular BHs as well as \whs\ with cosmological
  horizons --- for reviews see, e.g., \cite{viss-book,lobo,BR-book,BBS-17,WH-17} and references therein. 
  Our interest will be in \whs\ and regular BHs (BHs) without a center described by solutions of GR with a 
  minimally coupled scalar field as a source. Even such a narrow class of geometries turns out to be sufficiently 
  rich in properties of interest, especially concerning their dynamic stability.  
 
  A \wh\ is generally understood as a kind of tunnel or shortcut between two manifolds or two distant parts
  of the same manifold. However, this term is used in the literature in different meanings: the \whs\ can be 
  Lorentziian (traversable or not), Euclidean, and even quantum, which means that a wave function resembles 
  a tunnel geometry in a certain sense. In this paper, the term ``\wh'' is applied to traversable Lorentzian 
  \whs\ only. The so-called non-traversable \whs\ are, in general, BHs (or parts of BH space-times) rather 
  than \whs; thus, a \wh-like geometry quite usually appears as a spatial section of a \bh, and it is this
  phenomenon that was discovered by Flamm \cite{flamm} as early as in 1916 in his study of \Scz's solution,
  and his article is now referred to as the pioneering paper in \wh\ physics. As to Euclidean and quantum \whs,
  they represent quite separate areas of research.
     
  It is well known that the existence of traversable Lo\-rentz\-ian \whs\ as solutions to the Einstein equations
  requires what is called ``exotic matter'', i.e., matter violating the Null Energy Condition (NEC) 
  \cite{thorne, hoh-vis}, which is a part of the Weak Energy Condition (WEC) whose physical meaning is that
  the energy density is nonnegative in any reference frame. In particular, for \ssph\ systems sourced by scalar
  fields, \wh\ solutions can be and are really obtained if such a scalar field is phantom (or ghost), 
  i.e., has negative kinetic energy \cite{br73, ellis, vac1, SusZha}. In theories of gravity alternative to GR, 
  for example, scalar-tensor theories and the related multidimensional and curvature-nonlinear theories, 
  \wh\ solutions also appear only in the presence of phantom degrees of freedom 
   \cite{br73, bstar07,bstar09,bss10} (see also \cite{lobo, BR-book, WH-17} and references therein).
  This is true for both continuous matter distributions and thin shells \cite{bstar07}. In can happen that gravity
  itself becomes phantom in some region of space \cite{br73,br-JMP,bstar07}, in other cases the role of a phantom 
  is played by such geometric quantities as torsion \cite{almaz1,almaz2}, higher-dimensional metric components
  or variables related to higher-order derivatives \cite{EGB, HOG-13, b-kim03} or unusual couplings 
  between fields and matter \cite{Sush-Horn}. 
  For example, in brane-world theories, a source for \wh\ geometry in four dimensions can be provided by a
  tidal effect from extra dimensions originating from the Weyl tensor in the bulk \cite{b-kim03,BBS-17,kar-17}.
  Such a source, due to its geometric origin, is not subject to any energy conditions applicable to ordinary matter. 
  
  If, however, our interest is in obtaining (at least potentially) realistic \whs, it still makes sense to adhere to GR 
  and to use macroscopic matter or fields, because it is GR that explains all observations and experiments on 
  the macroscopic level; it is even used as a tool in applications like GPS navigation.

  Still, as yet nobody has observed macroscopic phantom matter, which puts to doubt its possible existence
  and therefore a possible realization of \whs, suitable, for instance, for interstellar communication and travel,
  even by any advanced civilization or in the remote future.
  
  In attempts to circumvent such problems and still to find \wh\ solutions in GR, a way of interest is to invoke
  such a sort of matter that would be phantom in a certain region of space only, somewhere in the vicinity
  of a \wh\ throat, while away from it it would observe all usual energy conditions \cite{trap1}.
  To obtain such a kind of matter, one can try to use a minimally coupled scalar field with the Lagrangian 
\beq                                                         \label{L_s}
      L_s =  h(\phi) g\MN \d_\mu \phi\d_\nu \phi - V(\phi),
\eeq
  where $h(\phi)$ and $V(\phi)$ are arbitrary functions. If $h(\phi)$ can change its sign, it cannot be absorbed 
  by redefinition of $\phi$ in its whole range. A situation of interest for us is if $h >0$ (so that the scalar field 
  is canonical and has positive kinetic energy) in a weak field region and $h < 0$ (a phantom, or ghost scalar
  field) in a strong-field region where one could expect a \wh\ throat. One can say that in this sense the ghost
  is trapped. Let us note that such a transition between $h > 0$ and $h < 0$ was considered in 
  \cite {rubin04} in a cosmological setting.

  It is known that phantom fields can produce not only \whs\ but also regular \bhs\ of different kinds, see, 
  e.g., \cite{pha1,pha2,BBS12}. Among such models, it makes sense to mark separately those combining 
  the features of BH physics and nonsingular cosmological models, the ones called {\it black universes\/}
  \cite{pha1,pha2}. Such objects look like ``conventional'' BHs (\sph\ ones in the existing examples) as seen
  from spatial infinity, where they can be \asflat, but after crossing the horizon, a possible explorer gets into 
  an expanding universe instead of a singularity. Thus such hypothetic objects combine the features of
  wormholes (no center but a regular minimum of the spherical radius $r(x)$), BHs  (static (R-) and nonstatic
  (T-) regions separated by a Killing horizon), and regular cosmological models. In addition, in such models the 
  \KS\ cosmology of the T-region can become isotropic at large times and be asymptotically de Sitter, making 
  these models a potentially viable description of an epoch before inflation. One can apply the trapped ghost 
  concept to such models on equal grounds with \whs\ \cite{trap2,trap3}: in such cases, the scalar field 
  should be phantom close to a minimum of the spherical radius $r$  (and this minimum can be 
  located both outside and inside the horizon or even coincide with it) but has canonical properties in the weak 
  field regions on both sides of the strong-field one, at large radii on the static side and at large times 
  on the cosmological side.

  It can also happen that a phantom field is not observed because it decays rapidly enough in the weak-field 
  region (the so-called ``invisible ghost'') but can also create \wh\ and regular BH geometries.

  In this paper we briefly review the \wh\ and black universe solutions of GR with minimally coupled 
  scalar fields, including trapped and invisible ghost fields, and also discuss the stability problem. 
  For any static model, the stability properties are of utmost importance since unstable objects cannot survive 
  in the real Universe, at least for a long time. It is known from previous studies that many \wh\ and black 
  universe solutions of GR are unstable under radial perturbations 
\cite{b-hod, stepan1,stepan2,hay,gonz1,gonz2,sta1,sta2}. 
  Considering the stability problem for trapped-ghost configurations, we shall see that it has some distinctive
  features that lead to a somewhat unexpected inference that transitions surfaces between canonical and 
  phantom regions of a scalar field play a stabilizing role. We will also discuss how the shape of the throat
  (its being generic or elongated) affects the stability study. 
  
  The paper is organized as follows. Section 2 presents the basic equations and some general features of 
  \sph\ \wh\ and regular BH space-times without a center. Section 3 describes the general properties of 
  space-times with scalar sources and presents a number of explicit solutions with ``simple'', trapped 
  and invisible ghosts. Section 4 discusses the stability problem for \sph\ scalar field configurations in GR
  and its particular features that emerge when we consider trapped-ghost scalars. Section 5 is a conclusion.

\section{Basic equations and general statements}

 We will restrict ourselves to considering only \ssph\ configurations and their small \sph\  perturbations.
 Before discussing solutions with scalar fields, let us begin with some general results which, using 
 spherical symmetry as the simplest illustration, reveal some general features of \wh\ solutions in GR.         

\subsection{General relations}  

 The general \ssph\ metric which can be written in the general form\footnote
      {Our conventions are: the metric signature $(+\ -\ -\ -)$, the curvature tensor
        $R^{\sigma}{}_{\mu\rho\nu} = \d_\nu\Gamma^{\sigma}_{\mu\rho}-\ldots,\
        R\mn = R^{\sigma}{}_{\mu\sigma\nu}$, so that the Ricci scalar
        $R > 0$ for de Sitter space-time and the matter-dominated
        cosmological epoch; the sign of $T\mN$ such that $T^0_0$ is the energy
        density, and the system of units $8\pi G = c = 1$.}
 without fixing the choice of the radial coordinate $u$:
\beq                                                           \label{ds}
        ds^2 = \e^{2\gamma(u)}dt^2 - \e^{2\alpha(u)}du^2 - \e^{2\beta(u)} d\Omega^2,
\eeq
  where $d\Omega^2 = d\theta^2 + \sin^2 \theta d\varphi^2$ is the linear element on a unit 
  sphere.\footnote
	{In what follows we use different radial coordinates, to be denoted for convenience
        by different letters:\\
	\cm $u$ --- a general notation,\\
	\cm $x$ --- a quasiglobal coordinate, such that $\alpha = -\gamma$,\\
	\cm $y$ --- a harmonic coordinate, such that $\alpha = 2\beta + \gamma$,\\
	\cm $z$ --- a ``tortoise'' coordinate, such that $\alpha = \gamma$.}
  Then the nonzero components of the Ricci tensor are 
\besub  \label{Ricci}
\bear 
       R^t_t \eql                                         \label{R00}
               -\e^{-2\alpha}[\gamma'' +\gamma'(\gamma'-\alpha'+2\beta')],
\yy 
      R^u_u \eql                           \label{R11}
         - \e^{-2\alpha}[\gamma''+2\beta'' +\gamma'{}^2+2\beta'{}^2-\alpha'(\gamma'+ 2\beta')],
\yy  
     R^\theta_\theta = R^\varphi_\varphi
		\eql \e^{-2\beta}                                    \label{R22}
               -\e^{-2\alpha}[\beta''+\beta'(\gamma'-\alpha'+2\beta')] ,
\ear
\esub
  where the prime stands for $d/du$. The Einstein equations can be written in two equivalent forms
\bearr                                                                   \label{EE}
    G\mN \equiv R\mN - \half \delta\mN R = - T\mN, \qquad {\rm or}
\qq
    R\mN = - (T\mN - \half \delta\mN T^\alpha_\alpha), 
\ear
  where $T\mN$ is the stress-energy tensor (SET) of matter.
  The most general SET compatible with the geometry \rf{ds} has the form 
\beq 		\label{SET}
	T\mN = \diag (\rho,\ -p_r,\ -p_T,\ -p_T),
\eeq
  where $\rho$ is the energy density, $p_r$ is the radial pressure, and $p_T$ is the 
  tangential pressure, which are in general different ($p_r \not\equiv p_T$), so that
  the SET \rf{SET} is anisotropic. It may contain contributions of one or several
  physical fields of different spins and masses or/and the density and pressures of one 
  or several fluids. 

  Our interest here is in the existence and properties of \wh\ and regular BH solutions to
  the Einstein equations. A \wh\ geometry with the metric \rf{ds} requires that the function
  $r(u) \equiv \e^{\beta(u)}$ should have a regular minimum $r = r_{\rm th}$ (this sphere 
  is called a throat) and reach values $r \gg r_{\rm th}$ on both sides of the throat. Of greatest
  interest are \wh\ geometries which are \asflat\ on one or both sides since only in this case 
  a \wh\ may be thought of as a local object in the modern, very weakly curved universe.  
  To distinguish \whs\ from BHs, it is often required that $g_{tt} \equiv \e^{2\gamma}$ should 
  be everywhere positive; however, it makes sense to admit $g_{tt} =0$ (horizons) sufficiently
  far from a throat, which may be of cosmological nature, with a possible de Sitter asymptotic beyond it. 

  As to regular BH geometries, among their different kinds \cite{pha2}, the most widely 
  discussed are those with a regular center, which can be obtained, for example, with a matter
  source satisfying the vacuum-like condition $\rho + p_r =0$, such as gauge-invariant nonlinear
  electrodynamics with Lagrangians of the form $L = L(f)$, $f \equiv F\mn F\MN$ ($F\mn$ being
  the Maxwell tensor) (see, e.g., \cite{dym-92, k-NED, ansoldi, BDG-12}). In the present paper we focus on 
  other kinds of regular \bhs, those which, like \whs, have no center, so that, in general, the spherical radius 
  $r = \e^\beta$ has a minimum. 

  Before considering such objects with scalar field sources, let us mention two general results concerning any 
  kinds of matter. To this end, let us use, for convenience, the so-called quasiglobal coordinate  $u = x$ under 
  the condition $\alpha+\gamma =0$; denoting
  $\e^{2\gamma} = \e^{-2\alpha} = A(x)$ and $\e^\beta = r(x)$, we rewrite the metric as
\beq          \label{ds-q}
           ds^2 = A(x) dt^2 - \frac{dx^2}{A(x)} - r^2 (x) d\Omega^2.  
\eeq
  The three different nontrivial components in the Einstein equations for the metric \rf{ds-q} have the form
\besub	\label{Gmn}
\bearr           											 \label{E00}
      G^t_t = \frac{1}{r^2}[-1 +A(2rr'' +r'^2) +A'rr'] = - T^t_t,
\\ \lal                                                                			\label{E11}
      G^x_x = \frac{1}{r^2}[-1 + A' rr' + Ar'^2] = - T^x_x,
\\ \lal						       					 \label{E22}
      G^\theta_\theta =  G^\phi_\phi = \frac{1}{2r}[2A r'' + rA'' + 2A'r'] = p_T,
\ear
\esub
  where the prime denotes $d/dx$, and \rf{E11} is the constraint equation, free from second-order derivatives.         

\subsection{The necessity of exotic matter}
    
  It is quite a well-known fact, first noticed for \ssph\ space-times \cite{thorne} and later proved for general
  static space-times in \cite{hoh-vis}. The term ``exotic matter'' is applied to matter whose SET violates the 
  Null Energy Condition (NEC) ($T\mN k^\mu k_\nu \geq 0$, where $k^\mu$ is any null vector, 
  $k^\mu k_\mu = 0$). This condition is, in turn, a part of the Weak Energy Condition (WEC) whose physical 
  meaning is that the energy density is nonnegative as viewed in any reference frame (see any textbook on GR). 
  
  The necessity of exotic matter for the existence of a \wh\ throat is easily shown 
  using \eqs \rf{E00} and \rf{E11} whose difference reads
\beq          \label{01q}
      2A\, r''/r = - (T^t_t - T^x_x) \equiv - (\rho + p_r),        
\eeq
  On the other hand, at a throat as a minimum of $r(x)$ we have
\beq
     r  > 0, \qquad       r' =0, \qquad         r''> 0.                   \label{min}
\eeq
  (In special cases where $r''=0$ at the minimum, it always happens that $r''> 0$ in its neighborhood.) 
  Then from \rf{01q} under the condition $g_{tt} = A>0$ it immediately follows  $\rho + p_r < 0$. 
  This inequality does indeed look exotic, but to see an exact result, we can choose the null vector
   $k^\mu = (1/\sqrt{A}, \sqrt{A}, 0, 0)$ and find that $T\mN k^\mu k_\nu = \rho+p_r$. Thus the 
   inequality $\rho + p_r < 0$ does indeed violate the NEC. 

  In the case of regular BHs it may happen that a minimum of $r(x)$ is located in a region 
  beyond its horizon, in which $A < 0$ (T-region), where the metric describes a \KS\ cosmology,
  In such a region, $x$ is a temporal coordinate, then $T^x_x = \tilde\rho$ is the energy density 
  while $ -T^t_t = p_t$ is the pressure in the (spatial) $t$ direction. Then the requirement 
  $r'' > 0$ leads, according to \rf{01q}, to $\tilde{\rho} + p_t < 0$. Thus such a minimum 
  also requires NEC violation. And lastly, if a minimum of $r$ coincides with a horizon, then the 
  same reasoning shows that NEC violation is necessary on either side in its neighborhood.  

\subsection{A no-go theorem for isotropic matter} 

  It is of interest whether or not \wh\ solutions can be obtained with a source in the
  form of isotropic matter (Pascal fluid), such that $p_r = p_T$. We will see that the answer
  is negative for \whs\ with flat or anti-de Sitter asymptotic behavior at both ends \cite{BBS-17}.

  If $p_r = p_T$, we have $G^x_x =G^\theta_\theta$, and the difference of \eqs \rf{E11}
  and \rf{E22} gives
\beq                                                   \label{iso1}
	r^2 A'' + 2A r r'' -2A r'^2 + 2 =0.
\eeq 
  The substitution $A(x) = D(x)/r^2(x)$ converts it to
\beq                                                   \label{iso2}
         D'' - \frac{4 D' r'}{r} + \frac{4 D r'^2}{r^2} + 2 =0.   
\eeq
  A possible minimum of $D(x)$ at some $x=x_0$ requires $D'=0$ and $D'' \geq 0$, and it should 
  be $D > 0$ for a traversable \wh. Meanwhile, if $D'= 0$,  \eqn{iso2} gives $D'' \leq -2$,
  hence a point where $D'=0$ is necessarily a maximum. 

  However, an \asflat\ traversable \wh\ requires $r\to \infty$ and $A \to 1$ as $x \to \pm\infty$, in an \asAdS\
  \wh\ it must be $A \sim r^2$ at large $r$, etc. In all such cases $D(x) \to \infty$ on both sides far from the 
  throat, hence it should have a minimum, which, as we have seen, is impossible. We thus have the following theorem:

\medskip\noi 
{\bf Theorem 1.} {\sl 
  A \ssph\ traversable wormhole with $r\to\infty$ and $A(x) r^2(x)\to \infty$  on both sides of 
  the throat cannot be supported by any  matter source with $p_r = p_T$ everywhere.    
 }

\medskip
  This excludes, in particular, twice \asflat\ and twice asymptotically AdS \whs\ as well as those \asflat\  on 
  one end and AdS on the other. What is not excluded, is that one or both asymptotic regions are de Sitter:
  in this case, $r \to \infty$ but $A \sim -r^2$ at large $r$, and it is not necessary to have a minimum of 
  $D(x)$. A number of examples of such \asdS\ \wh\  solutions have been found in \cite{BBS-17}, see also
  references therein.

  These inferences were obtained above using a specific coordinate condition, but they have an invariant 
  meaning since the quantities $A= g_{tt}$ and $r^2 = g_{\theta\theta}$ are insensitive to the choice of 
  the radial coordinate, as well as the mixed components $T\mN$ of the SET. 

  There exist \wh\ solutions with isotropic fluids as sources, but in all such cases the fluid occupies 
  a finite region of space, and there are inevitably ``heavy'' thin shells on the boundaries between 
  fluid and vacuum regions \cite{sush-05, lobo-05}. Such shells are highly anisotropic in the sense 
  that a tangential pressure is nonzero while the radial one is not defined (since the radial direction is 
  orthogonal to the shell). Therefore, these solutions do not contradict the above no-go theorem.

\section {Static systems with a scalar field source}

  The total action of GR with a minimally coupled scalar field $\phi$ as a source of gravity can 
  be written as 
\beq         \label{act}
             S = \frac{1}{2} \int \sqrt{-g} d^4 x
	 \Big[R + 2 h(\phi)  g^{\alpha \beta} \phi_{,\alpha} \phi_{,\beta} - 2V(\phi) \Big],
\eeq
  where, as before, $R$ is the scalar curvature, $g = \det(g_{\mu \nu})$,
  and $V(\phi^a)$ is a self-interaction potential. We include here an arbitrary function $h(\phi)$   
  and notice that $h(\phi) > 0$ for a normal scalar field with positive kinetic energy, and $h(\phi) < 0$ 
  for a phantom scalar. If $h(\phi)$ has the same sign in the whole range of $\phi$, it is 
  easy to redefine $\phi$ to obtain $h(\phi) = \pm 1$, but let us keep it arbitrary to be able to 
  consider solutions where $h(\phi)$ can change its sign. 
   
  The field equations may be written as
\bearr                                 							\label{eq-s}
	     2 h(\phi) \nabla^{\mu}\nabla_{\mu} \phi 
                       + \frac {dh}{d\phi}\phi^{,\mu}\phi_{,\mu}  + V_{\phi}= 0,
\yyy      				      								 \label{EE-s}
            R\mN = -2 h(\phi) \phi_{ , \mu} \phi_{ , \nu} + \delta\mN V(\phi)
\ear
  (recall that we are using the units in which $8 \pi G = 1$ and $c = 1$).  In \ssph\ space-time 
  with the metric \rf{ds}, assuming $\phi=\phi(u)$, the scalar field SET has the form
\bearr                                                              			 \label{SET-s}
     T\mN = h(\phi) \e^{-2\alpha} \phi'(u)^2 \diag (1,\ -1,\ 1,\ 1)
        + \delta\mN V(u).  			
\ear
  In terms of the metric \rf{ds-q} with the quasiglobal radial coordinate the field equations take the form
\besub   \label{EE-set}
\bear
     2(A r^2 h\phi')' - Ar^2 h'\phi' \eql r^2 dV/d\phi,  \label{e-phi}
\yy
              (A'r^2)' \eql - 2r^2 V;                         \label{00s}
\yy
               r''/r \eql - h(\phi){\phi'}^2 ;               \label{01s}
\yy
         A (r^2)'' - r^2 A'' \eql 2,                          \label{02s}
\yy                                                                \label{11s}
      -1 + A' rr' + Ar'^2 \eql r^2 (h A \phi'^2 -V),
\ear
\esub
  where the prime denotes $d/dx$. Equation \rf{e-phi} is the scalar field equation,  \rf{00s} is the component 
  $R^t_t = \ldots$, \rf{01s} and \rf{02s} are the combinations $R^t_t - R^x_x = \ldots$ and 
  $R^t_t - R^\theta_\theta = \ldots$, respectively, and \rf{11s} is 
  the constraint equation $G^x_x = \ldots$, free from second-order derivatives.

  We see that if $\phi \ne \const$, the SET \rf{SET-s} is anisotropic, and, in particular, Theorem 1 does not
  prevent the existence of twice \asflat\ \wh\ solutions. And indeed, such solutions are easily found with a
  massless phantom scalar field ($h <0, V\equiv 0$) \cite{br73, ellis}, see below.

  As to possible regular black hole solutions, there are significant restrictions, and let us consider them
  in some detail.

\subsection{Restrictions on \bhs\ with scalar fields}

\paragraph{Global structure theorem.} Equation \rf{02s} may be rewritten in terms of
   $B(x) \equiv  A(x)/r^2(x)$:
\beq 							\label{B''}
          r^4 B'' + 4 B' r r' + 2 = 0.
\eeq
  According to \eqn{B''}, if at some $x=x_0$ we have $B'=0$, then $B''(x_0) = -2/r^4(x_0) < 0$,
  so $x_0$ is a maximum of $B(x)$, and a regular minimum of this function is impossible. On the 
  other hand, since $r^2 >0$, regular zeros of $A(x)$, i.e., horizons, are also regular zeros of $B(x)$.  
  Since $B(x)$ has no minimum, this function, having once become negative while moving to the left 
  or to the right along the $x$ axis, cannot return to zero or positive values. Therefore, if $B(x) >0$ 
  in some range of $x$, it can have at most two zeros, and these zeros are simple since otherwise 
  there would be $B'=0$ and $B''>0$ near such a zero, contrary to \eqn{B''}. We obtain the 
  following theorem \cite{vac1}:

\medskip\noi
{\bf Theorem 2.} {\sl
	Consider solutions of \eqs \rf{EE-set}. Let there be a static region $a < x <b$ where 
	$a$ and $b$ may be finite or infinite. Then there are at most two horizons [$A(x) =0$], which 
  	are necessarily simple.
	}

\medskip
  By \eqn{B''}, a double horizon is also possible, but only if it separates two T-regions; in this case this 
  horizon is unique, and there is no static region at all.        
     
  All possible dispositions of zeros of the function $A(x)$, and hence the list of possible global 
  causal structures, turn out to be the same as for the vacuum solution with a cosmological constant,
  i.e., the \Scz-(anti-)de Sitter space-time. This conclusion is valid for any possible choice of the functions
  $V(\phi)$ and $h(\phi)$ since they are not involved in \eqn{B''}. The possible causal structures and
  the corresponding Carter-Penrose diagrams are listed, for example, in \cite{vac1, BR-book}.        

\paragraph{No-hair theorem.} 
  The expression ``BHs have no hair'' belonging to Wheeler \cite{MTW} means that BHs in GR are
  characterized by a restricted set of parameters (the mass, electric and magnetic charges and angular 
  momentum). There are a number of ``no-hair theorems'' claiming that no more charges or fields can
  accompany a BH under various circumstances. For us here it will be relevant to recall a theorem for the 
  \ssph\ system \rf{act} \cite{adler-78, bek-98} which can be formulated as follows in terms of the metric 
  \rf{ds-q} with the quasiglobal radial coordinate:

\medskip\noi {\bf Theorem 3.} {\sl
	Given \eqs \rf{EE-set} with $h(\phi) > 0$ and $V(\phi) \geq 0$, the only \asflat\ BH
       solution is characterized by $\phi = \const$ and the \Scz\ metric in the whole range    
       $x_h < x < \infty$, where $x=x_h$ is the horizon.
       }
       
\medskip
  Let us reproduce its proof following \cite{BR-book}. 

  With $h(\phi) > 0$, without loss of generality we can put $h(\phi) \equiv 1$ and also assume that 
  spatial infinity corresponds to $x\to \infty$. At the horizon $x=x_h$ we have by definition 
  $A = A(x_h) = 0$, and $A > 0$ at $x > x_h$. By Theorem 2, the horizon is simple, hence near 
  it $A \sim |x-x_h|$. Consider the function 
\beq                       \label{F}
              \cF(x) = \frac{r^2}{r'} (2V - A \phi'^2) 
\eeq
  One can verify that
\beq                          \label{Fx} 
		\cF'(x) = r \biggl( 4V + \frac{\phi'^2}{r'^2} + A \phi'^2 \biggr).
\eeq
  To do so, when calculating $\cF'$, one can substitute $\phi''$ from \rf{e-phi}, $r''$ from \rf{01s}, 
  and $A'$ from \rf{11s}. Let us integrate \rf{Fx} from $x_h$ to infinity:
\beq                                  \label{F-int}
		\cF(\infty) - \cF(x_h) = \int_{x_h}^\infty \cF'(x)\, dx.
\eeq
  Asymptotic flatness implies $r(x) \approx x$ at large $x$, therefore $r'(\infty) =1$, and $r'' \leq 0$ due 
  to \eqn{01s} with $h(\phi)>0$, so $r' > 1$ in the whole range of $x$, but $r'(x_h) < \infty$ (one can 
  verify \cite{BR-book} that $r' \to \infty$ would lead to a curvature singularity instead of a horizon). 

  The quantity $\cF(x_h)$ should be finite, since otherwise we would obtain infinite SET components and, 
  via the Einstein equations, a curvature singularity.

  If, however, we admit a nonzero value of $A\phi'^2$ at $x= x_h$, then, since $A=0$, it would mean 
  $\phi'^2 \sim (x-x_h)^{-1}$, and the integral in \rf{F-int} will diverge at $x = x_h$ due to the second term 
  in \rf{Fx}, which in turn leads to an infinite value of $\cF(x_h)$. Therefore $A\phi'^2 \to 0$ as $x\to x_h$,
  and we conclude that $\cF (x_h) = 2(r^2/r') V|_{x=x_h} >0$. On the other hand, $\cF(\infty) = 0$ due 
  to the asymptotic flatness condition. Thus, in \eq \rf{F-int} there is a nonpositive quantity in the left-hand 
  side and a nonnegative quantity on the right. The only way to satisfy \rf{F-int} is to put $V \equiv 0$ and
  $\phi' \equiv 0$ in the whole range $x>x_h$, and the only solution for the metric is then the \Scz\ solution
  with $r \equiv x$ and $A(x) = 1 - 2m/x$. 

  This concerned normal fields ($h(\phi) > 0$). The main consequence of Theorem 3 is that 
  {\sl nontrivial BH solutions with scalar ``hair'' require an at least partly negative potential $V(\phi)$.}

  Now, what happens if the scalar field is phantom, $h(\phi) < 0$?  
  It is straightforward to verify that the whole proof of the theorem can be preserved if we require 
  $V(\phi) \leq 0$. To prove that, it is sufficient to replace $\phi'^2 \mapsto -\phi'^2$ in all relations, then 
  $\cF$ and $\cF'$ will simply change their sign. The only subtle point is that now $r'' < 0$ due to \eqn{01s},
  therefore, to prove the theorem, we should separately require $r'(x_h) >0$. Thus {\sl nontrivial BH solutions
  with a phantom scalar field and $r'(x_h) >0$ require an at least partly positive potential $V(\phi)$.}

  To the author's knowledge, no similar theorem is known for scalar fields with $h(\phi)$ having an
  alternating sign, corresponding to the ``trapped ghost'' concept. We may expect, in particular, the 
  existence of BHs with such fields having completely positive or completely negative potentials. 
  
\subsection{Solutions with a massless scalar}

  After making clear the basic restrictions on possible solutions to \eqs\rf{EE-set}, let us begin a 
  consideration of their various examples with the simplest case, a massless scalar with $V(\phi) \equiv 0$. 
  Some properties of the solutions are immediately clear: thus, by Theorem 3, no \asflat\ BHs are possible
  if $h(\phi) >0$ or $h(\phi) < 0$, and  \wh\ solutions are impossible if $h(\phi) >0$.  As to $h(\phi)$ of 
  variable sign, some more reasoning is necessary. 

  For a massless field, the SET (\ref{SET}) with any $h(\phi)$ possesses the same structure as is known 
  for a usual massless scalar with $h = \pm 1$. Therefore for the metric we obtain the same well-known
  form as in these cases, which reduces to the Fisher metric \cite{fish} if $h(\phi) > 0$ and to its 
  counterpart for a phantom scalar, first found by Bergmann and Leipnik \cite{BerLei} (it is sometimes 
  called ``anti-Fisher'') if $h(\phi) < 0$. We here reproduce this solution in a joint form, following \cite{br73}. 
  To this end, it makes sense to return to the general metric \rf{ds}. 

  Two combinations of the Einstein equations \rf{EE-s} for the metric (\ref{ds}) and the SET (\ref{SET}) 
  with $V\equiv 0$ do not contain $\phi$ and read $R^t_t =0$ and $R^t_t + R^\theta_\theta =0$. 
  They can be most easily solved if we choose the harmonic radial coordinate $u = y$, defined by
  the condition $\alpha(y) = 2\beta(y) + \gamma(y)$. Indeed, the first of these equations takes the form
  $\gamma'' =0$, and the second one leads to the Liouville equation  
  $\beta'' + \gamma'' = \e^{2(\beta+\gamma)}$ (the prime here denotes $d/dy$). Their solution is
\bearr
     \gamma = - my,                                              \label{s}
\nnn
     \e^{-\beta-\gamma} =  s(k,y) := \vars {
                    k^{-1}\sinh ky,  \ & k > 0, \\
                                 y,  \ & k = 0, \\
                    k^{-1}\sin ky,   \ & k < 0.     }
\ear
  where $k$ and $m$ are integration constants, and other two constants have been excluded by choosing the zero
  point of the coordinate $y$ and the scale along the time axis. As a result, the metric takes the form \cite{br73}
\beq                                                         				  \label{ds1}
     ds^2 = \e^{-2my} dt^2 - \frac{\e^{2my}}{s^2(k,y)}
                    \biggr[\frac{dy^2}{s^2(k,y)} + d\Omega^2\biggl].
\eeq
  Note that without loss of generality we have $y \geq 0$, spatial infinity corresponds to $y = 0$, at small $y$ 
  the spherical radius $r$ behaves as $r \approx 1/y$, and $m$ has the meaning of the Schwarzschild mass. 

  All this was obtained from the general properties of the scalar field SET and does not depend on the choice
 of $h(\phi)$ in any way. Such a dependence emerges only when we take into account the constraint, i.e., 
  the ${1\choose 1}$ component of the Einstein equations (\ref{EE-s}) that leads to
\beq                                                           \label{int-0}
             k^2 \sign k = m^2 +  h(\phi) \phi'^2.
\eeq
  It means, in particular, that $h(\phi) \phi'^2 = \const$, hence $h(\phi)$ cannot change its sign as long 
  as we are dealing with a particular solution, characterized by fixed values of the integration constants $m$ 
  and $k$.\footnote
    {A detailed description of the properties of Fisher and anti-Fisher solutions 
     can be found in \cite{BR-book, cold08, SusZha, sigma}, see also references therein.
     Let us only mention here that the metric \rf{ds1} describes wormholes 
     \cite{br73, ellis} if the parameter $k$ is negative, which is only possible if $h (\phi) < 0$; 
     the two flat spatial infinities then correspond to $y=0$ and $y = \pi/|k|$.
    } 

  This situation does not change even if we use, instead of a single scalar field, a set of scalars $\phi^a$,
  forming  a nonlinear sigma model with the Lagrangian
\beq                                                           \label{L-sigma}
      L_\sigma = -h_{ab} g\MN \d_\mu \phi^a\d_\nu \phi^b,
\eeq
  where $h_{ab}$ are functions of $\phi^a$: in such a case, the metric has again the form (\ref{ds1}), and 
  instead of (\ref{int-0}) we have the relation \cite{sigma}
\[
      k^2 \sign k = m^2 +  h_{ab} (\phi^a)' (\phi^b)'.
\]
  Therefore the quantity $h_{ab} (\phi^a)' (\phi^b)'$ that determines the canonical or phantom nature
  of the scalar fields is inevitably constant. If the matrix $h_{ab}$ is positive-definite, the sigma model consists
  of canonical fields, and then $k > 0$, so \rf{ds1} is the Fisher metric containing a central naked singularity. 
  If, on the contrary, $h_{ab}$ is negative-definite, we are dealing with a set of phantom scalars, leading to
  solutions with $k < 0$ (wormholes), and, in addition, there is a subset of solutions with $k > 0$ containing
  horizons of infinite area which have been given the name of ``cold black holes'' \cite{cold08} because of
  their zero Hawking temperature. If the matrix $h_{ab}$ is neither positive- nor negative-definite, then
  there exist special solutions of \wh\ nature (with $k <0$) while other solutions correspond to a canonical
  scalar field and are described by Fisher's metric with a central singularity \cite{sigma}. However, there 
  cannot appear solutions of trapped-ghost nature. More complicated and more interesting examples can 
  appear only with nonzero potentials $V(\phi)$, such as those considered below.

\subsection {Scalar fields with a potential: Wormholes and black universes} 

  Let us return to field equations with an arbitrary potential, \eqs \rf{EE-set}, written for the metric \rf{ds-q} in terms
  of the quasiglobal coordinate $x$ (the ``gauge'' $\alpha + \gamma=0$ for the general metric \rf{ds}). It is hard 
  to obtain exact solutions with a prescribed potential $V(\phi)$, and, on the other hand, there is no clear physical 
  reason to prefer any specific potential if our purpose is to find solutions with physical properties of interest. Instead, 
  we will use the inverse problem method and choose the metric function $r(x)$ with required properties.

  It is easy to verify that \eqs \rf{e-phi} and \rf{11s} follow from (\ref{00s})--(\ref{02s}), which, if the potential
  $V(\phi)$ and the kinetic function $h(\phi)$ are known, form a determined set of equations for the unknowns
  $r(x)$, $A(x)$, $\phi(x)$. Furthermore, \eqn{02s} does not contain the scalar field, therefore, if $r(x)$ is known,
  then, solving \eqn{02s} to find $A(x)$, we determine the metric completely, after which $\phi(x)$ and $V(x)$ are
  found from \rf{01s} and \rf{00s}, respectively, and $V(\phi)$ is then defined unambiguously if $\phi(x)$ is monotonic. 
    
  Moreover, \eqn{02s} is easily integrated giving
\bear
            B'(x) \equiv (A/r^2)' = 2(3m - x)/r^4,          \label{B'}
\ear
  where (as before) $B(x) \equiv A/r^2,$ and the integration constant $m$ has the meaning of Schwarzschild mass
  if the metric (\ref{ds}) is \asflat\ as $x\to \infty$ (so that $r \approx x$, $A = 1 - 2m/x + o(1/x)$). If the metric
  is \asflat\ with $A\to 1$ as $x\to -\infty$, the Schwarzschild mass is there equal to $-m$ ($r \approx |u|$, 
  $A = 1 + 2m/|u| + o(1/u)$.

  This leads to a general result valid for {\it any\/} solution to \eqs \rf{EE-set} possessing two \asflat\ regions in the 
  presence of {\it any\/} potential  $V(\phi)$ compatible with such a behavior (such solutions can describe either \whs\ 
  or regular \bhs): the masses inevitably have opposite signs, as exemplified by the special case of a massless scalar ---
  the anti-Fisher \wh\ \cite{ellis, cold08, SusZha} whose metric in the gauge (\ref{ds-q}), easily obtained from \rf{ds1} 
  with $k < 0$, reads
\[
     ds^2 = -\e^{-2my} dt^2 + \e^{2my} [dx^2 + (k^2 + x^2) d\Omega^2],
\]
  where $y = |k|^{-1} \cot^{-1} (x/|k|)$ is the harmonic radial coordinate. 

  It is also clear that in solutions to \eqs \rf{EE-set} with $m = 0$ and an even function $r(x)$, the metric is 
  symmetric with respect to $x=0$, since $A(x)$ is also an even function according to \rf{B'}. Even is also $V(x)$
  found from \rf{00s}. However, the scalar field obtainable from \rf{01s} behaves in another way. 

  The simplest solution with a nonzero potential is obtained by putting $h(\phi) \equiv -1$ and \cite{pha1, pha2}
\beq                                                                                   \label{r1}
        r = (x^2 + a^2)^{1/2}, \cm a = \const > 0,
\eeq
  then \eqn {B'} leads to
\bearr                                                                               \label{B1}
      B(x) \equiv \frac{A(x)}{r^2(x)}
	      = \frac{c}{a^2} + \frac{1}{a^2+x^2} + \frac{3m}{a^3}
    			\left(\frac{ax}{a^2 + x^2} + \arctan \frac{x}{a}\right),
\ear
  with $c = \const$. Equations \rf{01s} and \rf{00s} then lead to expressions for $\phi(x)$ and $V(x)$:
\bear                                                        \label{phi1}
      \phi \eql \pm \arctan (x/a) + \phi_0,\qq		 \phi_0 = \const,
\\                                                                                   \label{V1}
     V \eql - \frac{c}{a^2} \frac{r^2 + 2x^2}{r^2} - \frac{3m}{a^3}
            \left( \frac{3ax}{r^2} + \frac{r^2 + 2x^2}{r^2}\arctan \frac{x}{a}\right)
\ear
  with $r = r(x)$ given by (\ref{r1}). In particular,
\beq                                                                                       \label{BV_as}
      B(\pm \infty) = -\frac 13 V(\pm \infty) = \frac{2ac \pm 3\pi m}{2a^3}.
\eeq
  Choosing in (\ref{phi1}), without loss of generality, the plus sign and
  $\phi_0=0$, we obtain for $V(\phi)$ 
\bearr   
     V(\phi) = -\frac{c}{a^2} (3 - 2\cos^2 \phi)             \label{Vf}
   		  - \frac{3m}{a^3} \big[3\sin\phi \cos\phi + \phi (3 - 2\cos^2\phi) \big].
\ear

  The solution behavior is controlled by two integration constants: $c$ that actually moves the plot of $B(x)$ up and
  down, and $m$ that affects the position of the maximum of $B(x)$. Both $B(x)$ and $V(x)$ are even functions
  if and only if $m = 0$, in agreement with the above-said. Asymptotic flatness at $x = +\infty$ implies
  $2ac = - 3\pi m$.

  Under this asymptotic flatness assumption, the $m =0$ solution describes the simplest symmetric configuration, 
  the Ellis \wh\ \cite{ellis}: $A \equiv 1$, $V \equiv 0$. At $m < 0$, by (\ref{BV_as}), we obtain a \wh\ with an AdS
  metric at the far end ($x\to -\infty$), corresponding to the cosmological constant $V_- < 0$. If  $m > 0$, so 
  that $V_{-} > 0$, there is a regular BH with a de Sitter asymptotic behavior far beyond the horizon, precisely
  corresponding to the above description of a black universe. These hypothetic configurations combine the
  properties of a wormhole (absence of a center, a regular minimum of the area function) and a BH (a Killing 
  horizon separating R- and T-regions).

  The horizon radius $r(x_h)$ can be obtained by solving the transcendental equation $A(x_h) =0$, where $A(x)$
  is given by \rf {B1}. It depends on the parameters $m$ and $a = \min r(x)$ and cannot be smaller than $a$, 
  which plays the role of a scalar charge since $ r \approx x$ and $\phi \approx \pi/2 - a/x$ at large $x$. Since
  $A(0) = 1 + c$, the point $x=0$ (minimum of $r$) is located in the static R-region if $c >-1$, i.e., if $3\pi m < 2a$
  (it is then a throat), precisely at the horizon if $3\pi m = 2a$, and in the T-region beyond the horizon if $3\pi m >2a$.
  This relationship between the parameters $m$ and $a$ prompt (and very probably it is the case in more general
  situations) that if the BH mass dominates over the scalar charge, then there is no throat in the static region, and
  a distant observer sees the BH approximately as usual in GR.

  As is clear from \eqs \rf{e-phi1} and \rf{00s}, the potential $V$ tends to a constant at each end of the $x$ range and, 
  moreover, we have there $dV/d\phi \to 0$. It is true for all classes of regular solutions mentioned in \cite{pha1}. More 
  precisely, {\it a regular scalar field configuration requires a potential with at least two zero-slope points (which are not 
  necessarily extrema) at different values of $\phi$}.

  Among suitable potentials are  $V = V_0 \cos^2 (\phi/\phi_0)$ and the Mexican hat potential 
  $V = (\lambda/4)(\phi^2 - \eta^2)^2$, with constants $V_0,\ \phi_0,\ \lambda,\ \eta$. It there is flat infinity at 
  $x = +\infty$, it certainly requires $V_+= 0$, while a de Sitter asymptotic can correspond to a maximum of $V$ since 
  phantom fields tend to climb up a slope of the potential instead of rolling down, as follows from \eqn {e-phi}. Note that 
  a consideration of spatially flat isotropic cosmologies with a phantom filed \cite{fara05} has shown that if $V(\phi)$ is 
  bounded above by $V_0 = \const > 0$, then the de Sitter solution is a global attractor. Quite probably, this result is
  also true for \KS\ cosmologies becoming isotropic at large times.

  The de Sitter expansion rate at late times is determined by the value of the potential $V_- >0$ (it corresponds to the 
  late-time effective cosmological constant value $\Lambda$) rather than by other details of the solution, such as the 
  Schwarzschild mass defined in the static region. A general conclusion is that black universes are a generic kind of solutions 
  to the Einstein-scalar equations  in the case of phantom scalars with proper potentials.

  The existence of black-universe solutions seems to be a natural feature of metric theories of gravity in the presence
  of any phantom degrees of freedom. All this leads to the idea \cite{pha1} that our Universe could emerge due to
  phantom-dominated collapse in some parent universe and undergo isotropization soon after crossing the horizon.
  It is known that \KS\ models of our Universe are not excluded by observations \cite{craw} if they became almost
  isotropic early enough, before the last scattering epoch (at redshifts $z\gtrsim 1000$). We are thus facing one more
  mechanism of universes multiplication, in addition to other known mechanisms such as, e.g., the chaotic inflation
  scenario.

\subsection{Models with a trapped ghost}

  The above example has shown that \wh\ and black-universe solutions must be quite a generic content of GR with
  a phantom scalar field. The trapped ghost concept potentially reconciles the existence of such objects with the 
  absence of phantom matter in a weak-field environment. 

  It is worth recalling that a variable sign of the kinetic term for a scalar field is not simply introduced {\it ad hoc\/} but 
  naturally follows from some models of multidimensional gravity, such as the one considered in \cite{kon-rub}, with the action
\beq                                     \label{act-D}
		S = \int d^D x \sqrt{g_D}\,[F(R) + c_1 R_{AB}R^{AB}+ c_2 R_{ABCD} R^{ABCD}]
\eeq 
  where $R$, $R_{AB}$ and $R_{ABCD}$ are the $D$-dimensional scalar curvature, Ricci and Riemann tensors, respectively,
  and $g_D = |\det g_{AB}|$. Under some additional assumptions, after reduction to a 4D theory and conversion to 
  Einstein's conformal frame one obtains the effective Einstein-scalar theory with the action 
\beq                                    \label{act-4E}
           S \sim \int d^4 x \sqrt{g_4} [R_4 + K_{\rm Ein}(\phi)(\d\phi)^2 - 2 V_{\rm Ein}(\phi)]
\eeq
  with certain functions $K_{\rm Ein}$ and $V_{\rm Ein}$ depending on the choice of \rf{act-D}. Choosing a quadratic
  function $f(R)$, for some particular parameter values, one obtains these functions in the form shown in Fig.\,1 \cite{kon-rub}. 

\begin{figure}
\centering
\includegraphics[width=6.5cm]{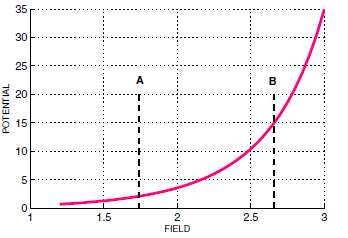}\quad
\includegraphics[width=6.5cm]{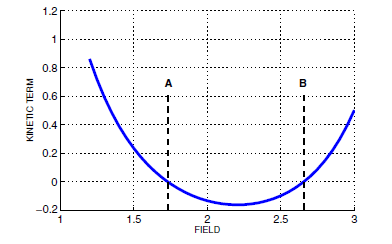}
\caption{\footnotesize
	The functions $V_{\rm Ein}(\phi)$ and $K_{\rm Ein}(\phi)$ for some parameters of
 	the action \rf{act-D}, showing a transition from canonical to phantom nature of the scalar field
       at values marked with the letters A and B.} 
\end{figure}

  Let us try to obtain particular solutions of this kind by choosing, as before, a proper shape of the function $r(x)$. 
  This function should possess the following properties:

\begin{enumerate}
\item
     For both \wh\ and black universes, a minimum of $r(x)$ must exist (located at $x=0$ without loss of generality), so that 
\[
        r(0) = a, \qquad r'(0) =0, \qquad r''(0) > 0, \qquad a=\const >0. 
\] 
\item
     By definition, in a trapped-ghost configuration it must be $h(\phi) < 0$ near a minimum of $r$, and $h(\phi) > 0$ far 
     from it. By (\ref{01s}), this implies $r'' > 0$ at small $|x|$ and $r'' < 0$ at sufficiently large $|x|$.
\item
     For obtaining \asflat\ or asymptotically (anti-) de Sitter models, we should have 
\[
        r(x) \approx |x|\qquad {\rm as}\qquad  x\to \pm \infty.
\]
\end{enumerate}

  A simple choice of the function $r(x)$ that satisfies the conditions 1--3 is \cite{trap3} 
  (see Fig.\,2)
\beq                      \label{r2}
         r(u) = a \frac{(x/a)^2+1} {\sqrt{(x/a)^2+n}},\cm   n = \const > 2,
\eeq
  where $a$ is is an arbitrary constant, related to the throat radius $r(0)$ by $r(0) = a/\sqrt{n}$, and the value of $a$ can 
  be used as a length scale. Note that the function \rf{r2} is different from $r(x)$ used in \cite{trap1,trap2}, and the 
  resulting expressions slightly simpler than there. 

  Further on we put $a = 1$; this actually means that the length scale remains arbitrary, but  $r$ and $m$ (it is the 
  Schwarzschild mass in our geometrized units) and other quantities with the dimension of length are expressed in units of $a$,
  accordingly, $B$, $V$ and other quantities with the dimension $\rm (length)^{-2}$ are expressed in units of $a^{-2}$, etc.; 
  the quantities $A, \phi, h$ are dimensionless. Since
\beq                          \label{r2''}
       r''(x) = \frac{x^2(2-n) + n(2n-1)}{(x^2 + n)^{5/2}},
\eeq
  we obtain $r'' > 0$ at $x^2 < n(2n-1)/(n-2)$ and $r'' < 0$ at larger $|x|$, as required; it is also clear that $r \approx |x|$ 
  at large $|x|$. It guarantees $h < 0$ at small $|x|$ and $h > 0$ at large $|x|$ (see Fig.\,2, right panel). 
\begin{figure}
\centering
\includegraphics[width=6.5cm]{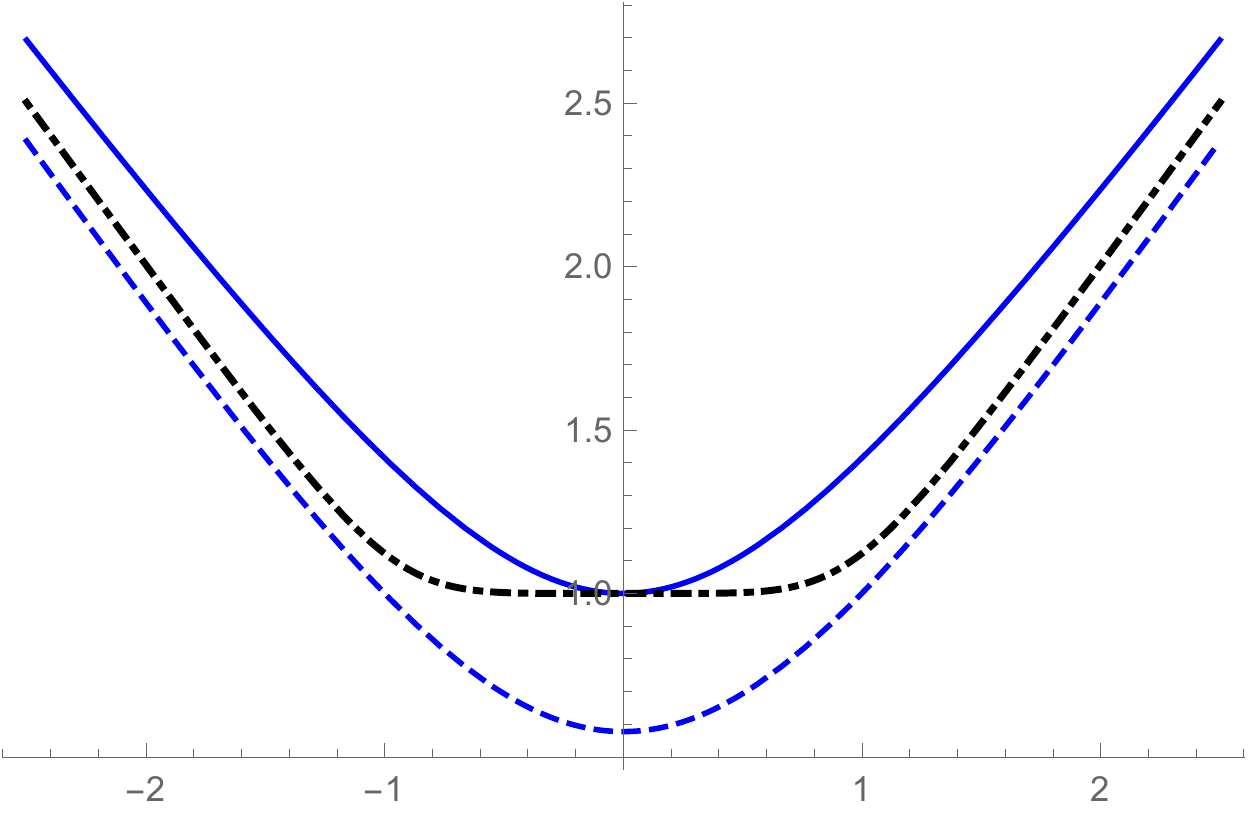}\
\includegraphics[width=6.5cm]{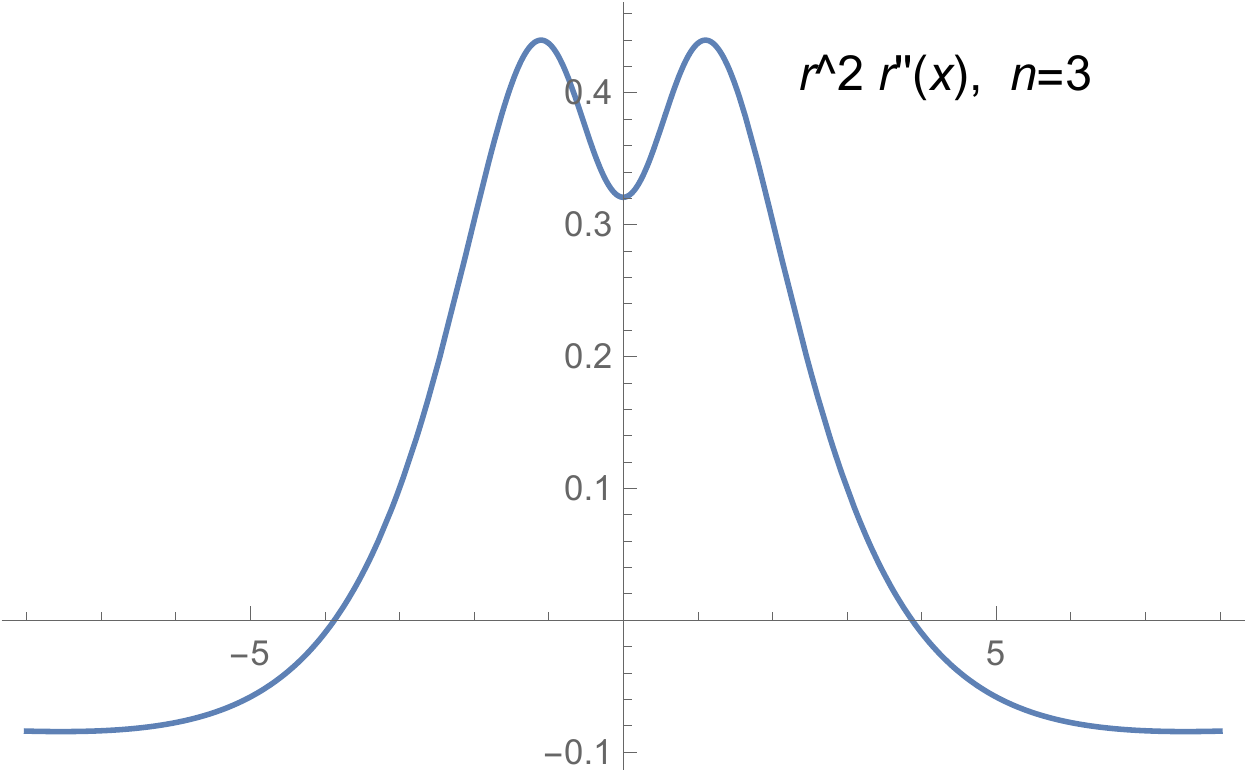}
\caption{\footnotesize
	Left: Plots of $r(x)$ under different assumptions: by \rf{r1} (solid), by \rf{r2}, $n=3$ (trapped ghost, dashed),
	and by \rf{r3} (long throat, dot-dashed). Right: Plot of $r^2 r''$ for \rf{r2}, $n=3$; zeros
      of this function show the surfaces $h(\phi) =0.$ }
\end{figure}
  
  To avoid cumbersome expressions for $B(x)$ and other quantities, let us restrict our discussion to the value $n = 3$. An 
  inspection shows that a particular choice of the parameter $n > 2$ does not change the qualitative features of the solutions.

  Integrating \rf{B'} for $n=3$, we obtain
\beq                              \label{B2}
        B = B_0 + \frac{26 + 24 x^2 + 6 x^4 + 3mx (69 + 100 x^2 + 39 x^4)}{6(1+x^2)^3}
        + \frac{39 m}{2} \arctan x,
\eeq
  where $B_0$ is an integration constant.

  Now suppose that our system is \asflat\ as $x \to + \infty$. Since $B=A/r^2$ and $A\to 1$ at infinity, we require
  $B\to 0$ as $x\to \infty$ and thus fix $B_0$ as
\beq                                                        \label{B02}
		    B_0 = -\frac{39 \pi m}{4}.
\eeq
  The form of $B(x)$ (and accordingly $A(x) = Br^2$) substantially depends on the mass $m$,  see Fig.\,3. It is clear that
  with $m <0$ we obtain $B(x)$ that tends to a positive constant as $x \to -\infty$, so that $A \sim r^2$, and a \wh\ with 
  an AdS asymptotic behavior at the far end is obtained (an M-AdS \wh\ for short, where M stands for Minkowski). 
  If $m=0$, we obtain a twice  \asflat\ (M-M) \wh. Lastly, if $m > 0$, then $B(x)$ changes its sign at some $x$ and tends to
  a negative constant, which means that there is a black universe tending to de Sitter geometry as $x \to -\infty$, in full
  analogy with our first example \rf{r1}--\rf{Vf}.  

  Now the metric is known completely, while $\phi(x)$ and $V(\phi(x))$ are found, as before, from \eqs (\ref{01s}) and 
  (\ref{00s}). To construct $V$ as an unambiguous function of $\phi$ and to find $h(\phi)$, it makes sense to use the 
  parametrization freedom for $\phi(x)$ and to choose 
\beq                             \label{phi_x2}
        \phi(x) = \frac{1}{\sqrt{3}}\arctan\frac{x}{\sqrt{3}},
\eeq
  a behavior common to kink configurations, such that $\phi$ has a finite range: $\phi \in (- \phi_0, \phi_0)$,
  $\phi_0 = \pi/(2\sqrt{3})$. Thus we have $x = \sqrt{3}\tan(\sqrt{3}\phi)$, ant its substitution to the expression for 
  $V(x)$ found from (\ref{00s}) gives $V(\phi)$ defined in this finite range. 

  The kinetic coupling function $h(\phi)$ is then expressed from (\ref{01s}):
\beq
        h(\phi) = \frac{x^2-15} {x^2+1}                  \label{h2}
        = \frac{3\tan^2(\sqrt{3}\phi)-15}{3\tan^2(\sqrt{3}\phi)+1}.
\eeq
  This function is also defined on the interval $(-\phi_0,\phi_0)$, which may be extended to $\R$ if we suppose 
  $h(\phi)\equiv 1$ at $|\phi|\geq \phi_0$. As is clear, the NEC is violated where and only where $h(\phi) < 0$. 
\begin{figure}
\centering
\includegraphics[width=5.5cm]{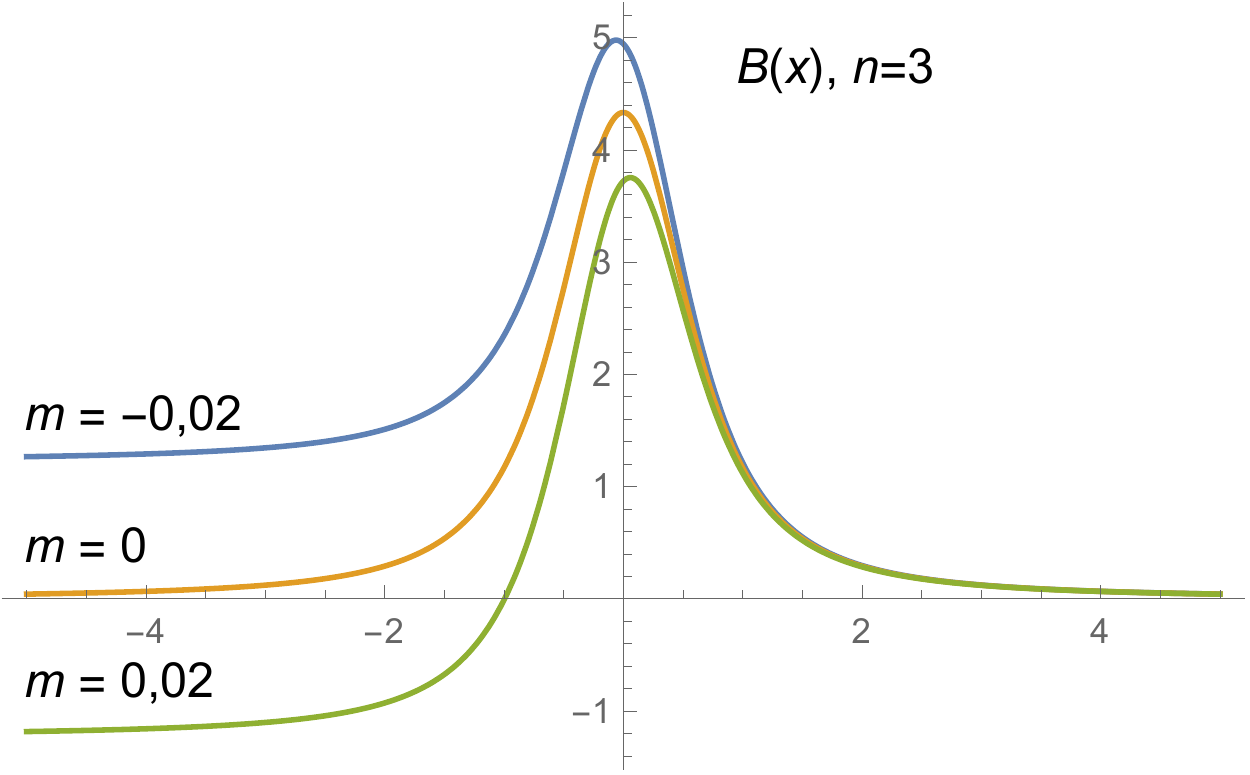}
\includegraphics[width=5.5cm]{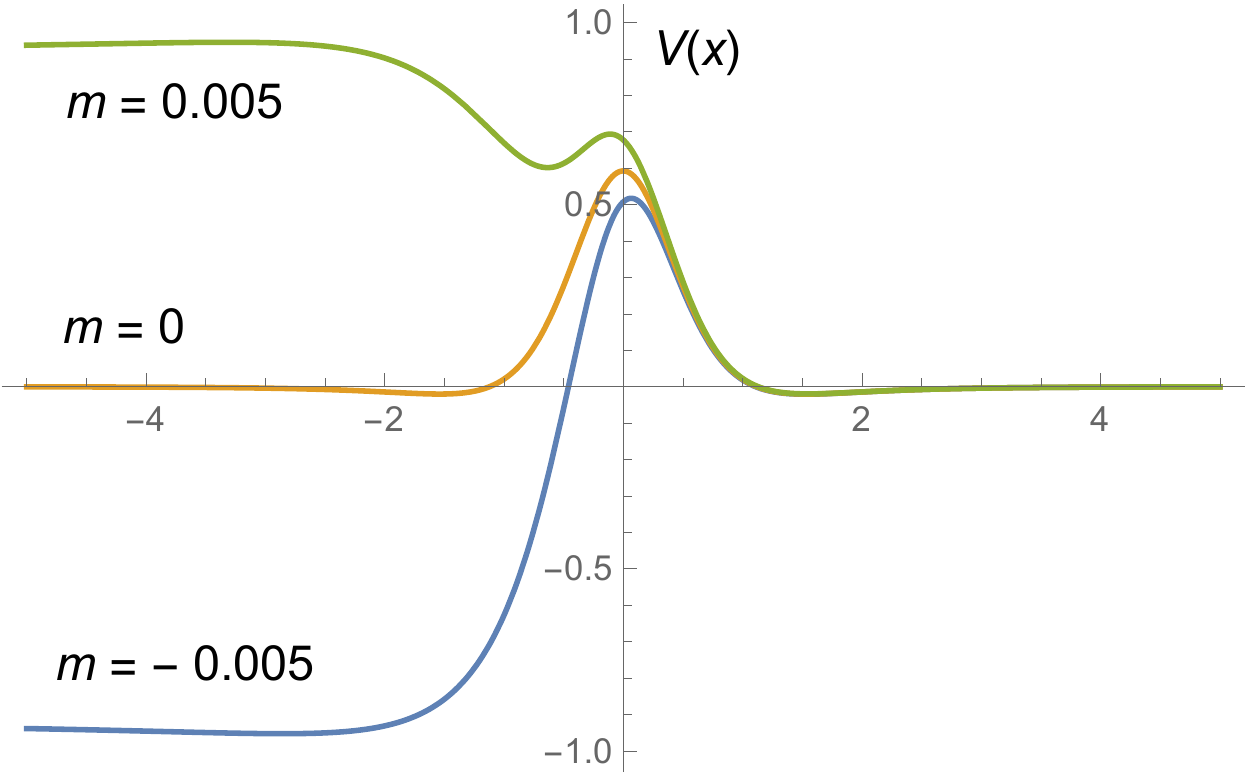}
\includegraphics[width=3cm]{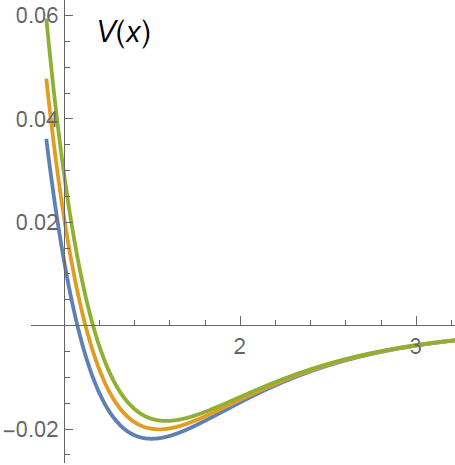}
\caption{\footnotesize
     Plots of $B(x)$ (left) and $V(x)$ (middle) for different values of $m$ in a model with $r(x)$ given by \eqn{r2}
     with $n=3$.  With $m < 0$ we obtain an asymmetric M-AdS \wh, with $m=0$ a symmetric M-M \wh,  and
     with $m > 0$ a black universe with de Sitter behavior as $x \to -\infty$. The right panel shows the behavior
     of $V$ at larger $x$, where the curves almost merge. (The colors there have the same meaning as in other panels.)}
\end{figure} 

  For $V(x)$ we obtain \cite{trap3}
\bearr
        V(x) =  \frac 1 {12 (1 + x^2)^2 (3 + x^2)^3}\Big\{ 32 (-6 + x^2 + 3 x^4) 
\nnn \inch
          +  6 m x (2655 + 6930 x^2 + 5420 x^4 + 1326 x^6 + 117 x^8) 
\nnn \inch
         + 117 m (1 + x^2)^2 (15 + 89 x^2 + 29 x^4 + 3 x^6) (-\pi + 2 \arctan x)\Big\}.
\ear
  
  The function $V(\phi)$ can also be extended to the whole real axis, $\phi \in \R$, by putting $V(\phi)\equiv 0$ at all 
  $\phi \geq \phi_0$ (since $V(x=+\infty)=0$) and $V(\phi) = V(-\phi_0) > 0$ at all $\phi < -\phi_0$.

  One can easily verify that the values of $B(x)$ as $x\to -\infty$ are directly related to the asymptotic values of the 
  potential $V$, which plays the role of an effective cosmological constant at large negative $x$:
\beq
             V(-\infty) = 117 m \pi/2 = -3 B(-\infty).                       \label{as-V}
\eeq
  We see that the trapped-ghost solution preserves all qualitative features of the simplest solution with the 
  dependence \rf{r1} of the spherical radius. 

  Other, more complicated solutions with diverse global structures are obtained if, besides a scalar field, we introduce an 
  electromagnetic field. Such solutions, including M-M and M-AdS wormholes as well as regular black holes having up to 
   three horizons (up to four horizons if asymptotic flatness is not required), have been found in \cite{trap3}; their structures 
  turn out to be quite similar to those obtained earlier with a purely phantom scalar field \cite{BBS12}.  

\subsection {Wormholes with an invisible ghost and a long throat}

  In the previous subsection, having admitted the existence of phantom fields, we discussed a way to explain why they are 
  not observed under usual conditions using the ``trapped ghosts'' concept. Another way to explain the same is what may 
  be called the ``invisible ghost'' concept, which means that the phantom field decays rapidly enough at infinity and is there 
  too weak to be observed \cite{b-kor1, invis2}. To achieve this goal, we need rapidly decaying quantities $\phi'$ and hence, 
  by \eqn{01s} $r''/r$. Let us therefore replace the previous ansatz \rf{r1}, $r = (a^2+x^2)^{1/2}$, with \cite{invis2}
\beq      \label{r3}
	 r(x) = a (1 + x^{2n})^{1/(2n)}, \cm n \in \N,
\eeq
  where $a > 0$ is, as before, an arbitrary constant, now equal to the throat radius. We will again put $a = 1$, so that lengths 
  are expressed in units of the throat radius; the quantities like $B(x)$ and $V(x)$ with the dimension (length)$^{-2}$ are 
  expressed in units of $a^{-2}$, while the dimensionless quantities $A(x)$ and $\phi(x)$ are insensitive to this assumption.

  The value $n=1$ returns us to the ansatz \rf{r1}. Higher values of $n$ lead to a new feature of the space-time geometry: 
  the spherical radius $r(x)$ is changing quite slowly near the throat $x=0$, making it possible to call it {\it a long throat}, 
  see the dot-dashed curve in Fig.\,2, left panel. At large $|x|$ we now have $r''/r \approx (2n-1) x^{-2n-2}$, hence 
  $\phi' \sim 1/x^{n+1}$, which at large enough $n$ conforms to the ``invisible ghost'' concept.

\begin{figure*}
\centering
\includegraphics[width=5.5cm,height=5.5cm]{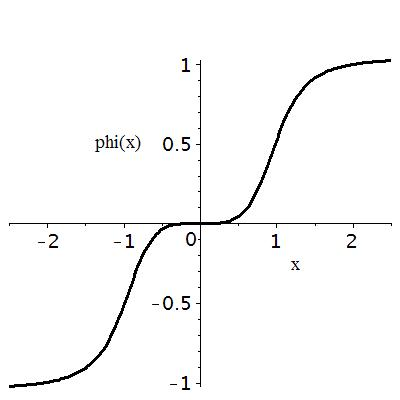}
\qquad
\includegraphics[width=5.5cm,height=5.5cm]{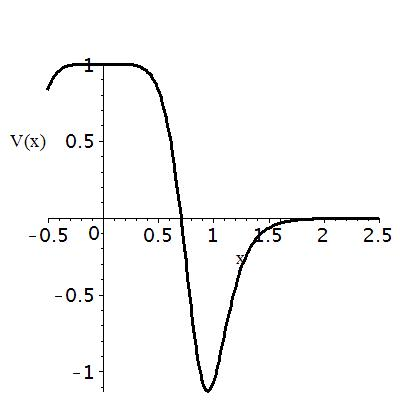}
\caption{\footnotesize
	The scalar field $\phi(x)$ (left) and the potential $V(x)$ (right) for $n=4$.}
\end{figure*}

  Let us put $m=0$, restricting ourselves to massless \whs. Then $B'(x)$ (\ref{B'}) is an odd function:
\beq                     \label{B'3}
	 B'(x) = - \frac{2x}{(x^{2n} + 1)^{2/n}},
\eeq
  whose integration gives
\beq 		 \label{B3}
     B(x) =  -x^2  F \Big( \frac{1}{n}, \frac{2}{n};1 + \frac{1}{n}; -x^{2n} \Big) + B_0,
\eeq
  where $B_0$ is an integration constant, and $F(a, b; c, z)$ is the Gaussian hypergeometric function. Assuming asymptotic 
  flatness at large positive $x$, since $B = A/r^2$ and $A \to 1$ at infinity, we require $B \to 0$ as $x \to\infty$ and thus 
  fix $B_0$ as
\beq  			 \label{B0}
       B_0 = \lim\limits_{x\to \infty} x^2 F \Big( \frac{1}{n}, \frac{2}{n};1 + \frac{1}{n}; -x^{2n}\Big)
\eeq

  We see that it is a twice asymptotically flat (M-M) wormhole, and a plot of $B(x)$ is similar to the curve $m=0$ in Fig.\,3 (left).
  Curiously, the behavior of $A(x)$ shows that there is a domain of repulsive gravity around the throat. 

  Now the metric is known completely, while $\phi(x)$ and $V(\phi(x))$ are again easily found from \eqs \rf{01s} and \rf{00s}.
  The expression for the scalar field $\phi(x)$ in the case $n=4$ is (assuming $\phi(0)=0$)
\beq 		 \label{phi-3}
              \phi(x) =  \frac{\sqrt{7}}{4} (\sign x) \arctan (x^4),
\eeq
  see Fig.\,4 (left). For the potential $V(x)$ there is rather a cumbersome expression in terms of hypergeometric functions, 
  gamma functions and Legendre functions, and we will not present it here. It is plotted in Fig.\,4 (right). Since $V(x)$ is an 
  even function, the plot is restricted to $x \geq 0$. 

  Models with diverse global structures emerge if there is a nonzero Schwarzschild mass $m$ or/and, besides a scalar field, we 
  consider an electromagnetic field with the corresponding electric or magnetic charge. Examples of such solutions, including
  M-M, M-dS (de-Sitter), M-AdS (anti de-Sitter) wormholes and regular black holes containing up to four horizons, can be found
  in \cite{BBS12,trap2,trap3,b-kor1}. The global qualitative features of our present field system are similar to those described 
  above, and the same kinds of regular solutions can be obtained using the same methods. 

\section{The stability problem}
 
\subsection{Spherically symmetric perturbation equations}

  The stability problem is of great importance while studying any equilibrium configurations since only stable or very slowly 
  decaying ones have a chance to exist for a sufficiently long time. On the other hand, the development of instabilities can lead 
  to a lot of important phenomena, such as, for example, structure formation in the early Universe and Supernova explosions. 

  Let us discuss the stability problem for configurations with scalar fields like those described in the previous section. The problem 
  is whether an initial small time-dependent perturbation can grow strongly enough to destroy the system. As in a majority of
  such studies, we consider linear perturbations and neglect their quadratic and higher-order combinations. Moreover, we restrict
  the study to perturbations that preserve spherical symmetry (in other words, only radial, or monopole perturbations). On one 
  hand, they are the simplest, but on the other, they are the most ``dangerous'' ones since they lead to instabilities in many 
  known models with self-gravitating scalar fields \cite{b-hod, stepan1, stepan2, hay, gonz1, gonz2, sta1, sta2}. 
  Other modes of perturbations are usually found to be stable, see, e.g., \cite{sta2, b-sha13}. Let us suppose that some static 
  solution is already known (the solutions described above being special cases) and consider its time-dependent perturbations.  
   
  We will follow the lines of  \cite{BR-book, sta1, b-kor1, in-lobo} and consider the same field 
  system as before, that is, \rf{act}, \rf{eq-s}, \rf{EE-s}, with the scalar field SET:
\beq
      T\mN [\phi] = h(\phi)[ 2\phi_\mu \phi^\nu - \delta\mN \phi^{\alpha}\phi_{\alpha}] + \delta\mN V(\phi),
\eeq
  The general \sph\ metric can be written in the form \rf{ds}, that is,
\beq                                                                                                    \label{ds4}
    ds^2 = \e^{2\gamma} dt^2 - \e^{2\alpha}du^2 - \e^{2\beta}d\Omega^2,
\eeq
  where now $\gamma$, $\alpha$ and $\beta$ are functions of both the radial coordinate $u$ and time $t$. 
  We use again the notation $r(u) \equiv \e^\beta$ and preserve the freedom of choosing the coordinate $u$.

  Consider linear \sph\ perturbations of static solutions (known by assumption) to the field equations due to (\ref{act}).
  Now we write
\beq
          \phi (u,t) = \phi(u) + \delta\phi(u,t), \qquad \gamma(u,t) = \gamma(u) + \delta\gamma
\eeq
  for the scalar field and the metric function $gamma$ and similarly for all other quantities, assuming small ``deltas''. 
  
  All nonzero components of the Ricci tensor are (with only linear terms with respect to time derivatives) 
\bear
     R^t_t \eql                                               \label{R00t}
     \e^{-2\gamma}(\ddot\alpha + 2\ddot\beta)
           -\e^{-2\alpha}[\gamma'' +\gamma'(\gamma'-\alpha'+2\beta')],
\yy                                                            \label{R11t}
     R^u_u \eql 
     \e^{-2\gamma}\ddot\alpha                           
     - \e^{-2\alpha}[\gamma''+2\beta'' +\gamma'{}^2
     +2\beta'{}^2 - \alpha'(\gamma'+2\beta')],
\yy                                                                                                   \label{R22t}
     R^\theta_\theta \eql R^3_3 = \e^{-2\beta} +\e^{-2\gamma}\ddot\beta
              -\e^{-2\alpha}[\beta''+\beta'(\gamma'-\alpha'+2\beta')],
\yy
     R_{tu}\eql  2[\dot\beta' + \dot{\beta}\beta'                                      \label{R01t}
                 -\dot{\alpha}\beta'-\dot{\beta}\gamma'],
\ear
  where dots and primes denote $\d/\d t$ and $\d/\d u$, respectively.

  The background (zero-order, static) scalar field equation and the ${t\choose t}$, 
  ${u\choose u}$, and ${\theta\choose \theta}$ components of the Einstein equations (\ref{EE}) read
\bear                                                  \label{e-phi0}
     2h[\phi'' + \phi'(\gamma'+2\beta'-\alpha')]
     		+ h'\phi' \eql \e^{2\alpha}dV/d\phi;
\yy                                                     \label{00-0}
     \gamma'' + \gamma'(\gamma'+2\beta'-\alpha') \eql -\e^{2\alpha} V;
\yy                                                         \label{11-0}
     \gamma'' + 2\beta'' + \gamma'{}^2 + 2\beta'{}^2
            -\alpha'(\gamma'+2\beta') \eql  -2h \phi'^2 - \e^{2\alpha}V ;
\yy                                                        \label{22-0}
     -\e^{2\alpha-2\beta}
       + \beta'' + \beta'(\gamma'+2\beta'-\alpha') \eql -V\e^{2\alpha}.
\ear
   The first-order perturbed equations (scalar, $R_{ut}=\ldots$, and $R^\theta_\theta = \ldots$) have the form
\bearr                                                     \label{e-phi1}
     2\e^{2\alpha-2\gamma} h \delta\ddot\phi - 2h[\df''
                     +\df' (\gamma'+2\beta'-\alpha')	+\phi'(\dg' + 2\db' -\da')]	
\nnn \inch
        - 2\delta h[\phi'' + \phi'(2\beta'+\gamma'-\alpha')]
	- h' \df' - \phi'\delta h' +  \delta(\e^{2\alpha}V_\phi) =0,
\yyy                                                       \label{01-1}
    \delta\dot\beta' + \beta'\delta\dot\beta
        - \beta' \delta\dot\alpha - \gamma' \delta\dot{\beta}
                = - h \phi'\delta\dot\phi,
\yyy                                                       \label{22-1}
    \delta(\e^{2\alpha-2\beta})
            + \e^{2\alpha-2\gamma} \delta\ddot\beta
                    -\db'' -\db'(\gamma'+2\beta'-\alpha')
\nnn \inch\cm
        -\beta'(\dg' + 2\db' -\da')  = \delta(\e^{2\alpha} V).
\ear
   Equation (\ref{01-1}) is easily integrated in $t$; being interested in time-dependent perturbations only, we omit the arbitrary
   function of $u$ that emerges at this integration and describes static perturbations. This leads to
\beq                                                       \label{01-1i}
              \db' + \db(\beta'-\gamma') - \beta'\da = -h \phi' \df.
\eeq

  Our system possesses two independent forms of arbitrariness: the freedom to choose a {\it radial coordinate\/} $u$ in the 
  static background solution, and the {\it perturbation gauge\/}, a freedom related to the choice of a reference frame in
  perturbed space-time. As a result, we can impose some relation containing $\da,\ \db$, etc. Further on we employ both 
  kinds of freedom. All the above equations are written in a universal form, where neither the $u$ coordinate nor the 
  perturbation gauge are fixed.

  Let us simplify the equations by choosing the ``tortoise'' coordinate $u=z$, specified by the condition $\alpha = \gamma$
  (it is the best for wave equations), and the perturbation gauge $\db \equiv 0$. Then from \eqn{01-1i} we find $\da$ in 
  terms of $\df$ (now the prime stands for $d/dz$):
\beq                                  			      \label{da-df}
       \beta' \da = h(\phi) \phi' \df.
\eeq
  From \eq (\ref{22-1}) we find $\dg' - \da'$ in terms of $\da$ and $\df$:
\beq                                  \label{dg-df}
    \beta'(\dg'-\da') = 2 \e^{2\alpha-2\beta}\da - \delta(\e^{2\alpha} V).
\eeq
  With all that substituted to (\ref{e-phi1}), the following wave equation is obtained:
\bearr                                                        \label{eq-df}
        \delta\ddot\phi  -\df'' - \df' (2\beta' + h'/h)+ U \df =0,
\yyy                                                          \label{def-U}
        U \equiv \e^{2\alpha}\biggl[\frac{2h \phi'^2}{\beta'^2}(V - \e^{-2\beta})
          + \frac{2\phi'}{\beta'} V_\phi + \frac{V_{\phi\phi}}{2h}\biggr]
          -\frac{h'' + h_\phi \phi''}{2h} -\frac{2\beta' h'}{h},\cm
\ear
  where the index $\phi$ denotes $d/d\phi$. The expression \rf{def-U} for $U$ is a generalization of 
  the one obtained in \cite{sta1} for scalar-vacuum systems with $h = \ep =\pm 1$.

  Next, the first-order derivative of $\df$ in \rf{eq-df} is removed by the substitution 
\beq                                                  	     \label{to_psi}
       \df = \psi(z,t) \e^{-\eta}, \cm \eta' = \beta' + \frac{h'}{2h},
\eeq
  which reduces the wave equation to the canonical form
\beq                                                        \label{wave}
       \ddot \psi - \psi'' + \Veff (z)\psi =0,
\eeq
  with the effective potential
\bearr                                                     \label{Veff}
     \Veff (z) = U + \eta''+ \eta'{}^2
            = \e^{2\alpha}\biggl[\frac{2h \phi'^2}{\beta'^2}(V - \e^{-2\beta})
                       + \frac{2\phi'}{\beta'} V_\phi 
		- \frac{h_\phi}{4h^2} V_\phi + \frac{V_{\phi\phi}}{2h}\biggr]
		+ \beta'' + \beta'^2.\qquad
\ear
  One more substitution, using the static nature of the background,
\beq                                                       \label{to_y}
       \psi (x, t) = Y(x) \e^{i\omega t}, \cm \omega = \const,
\eeq
   leads to the \Schr-like equation
\beq                                                        \label{Schr}
      Y'' + [\omega^2 - \Veff(x)] Y =0.
\eeq
  
  Now, if there exists a nontrivial solution to \rf{Schr} such that $\im \omega < 0$, for which some physically reasonable 
  conditions hold at the ends of the range of $z$ (including the absence of ingoing waves), then we can conclude that the 
  static background system is unstable since the perturbation $\df$ can exponentially grow with time.  Otherwise our static 
  system is stable in the linear approximation. The value $\omega=0$ makes possible a linear growth of perturbations.
  As usual in such studies, the stability problem is thus reduced to a boundary-value problem for 
  \eq (\ref{Schr}) --- see, e.g.,  \cite{BR-book, sta1, sta2, b-hod, cold, stepan1, stepan2, gonz1, b-sha13}.

  The gauge $\db = 0$ is convenient for calculations but looks doubtful when applied to configurations with throats. The
  reason is that writing $\db = 0$, we suppose that the throat radius is not subject to perturbations, whereas its changes 
  are in general admissible \cite{cold, gonz1, BR-book}. It might seem that the pole in $\Veff$ related to $\db$ in the denominator 
  in (\ref{def-U}) is an artifact of this gauge. It turns out, however, in full similarity with \cite{gonz1, sta1, BR-book}, that
  \eqn{Schr} is in fact gauge-invariant, and the perturbation $\df$ represents a certain gauge-invariant quantity in the gauge 
  $\db =0$. This issue is discussed in detail, e.g., in \cite{sta1,BR-book,in-lobo}.

\subsection {Perturbations near a generic throat}
  
  Since \eqn {Schr} is gauge-invariant, boundary value problems in the stability study are really meaningful. However, the 
  problems are rather complicated due to the singular nature of the effective potential $\Veff$. The singularities are of two 
  kinds: one is related to throats, if any, due to terms containing $1/\beta'^2$ and $1/\beta'$ in $\Veff$ (since $\beta'=0$ 
  on a throat); the other takes place at transition surfaces from usual to phantom scalar fields because $\Veff$ contains terms
  proportional to $h^{-2}$ and $h^{-1}$, while $h=0$ at such a transition. Let us first discuss the singularities related to throats, 
  following \cite{in-lobo} and partly \cite{gonz1,sta1}. 

  Suppose that there is a throat at (say) $z=0$, where the spherical radius $r(z) = \e^\beta$ possesses a generic minimum, 
  such that  $r(z) = r_0 + \half r_2 z^2  + o(z^2)$, where $r_0 >0,\ r_2>0$. Also, without loss of generality, close to the 
  throat we put $h(\phi) = -1$ since due to \rf{01s} it should be $h <0$, and $h(\phi) = -1$ is obtained by a regular 
  redefinition of $\phi$. Then, from the zero-order (background) equations it follows that near the throat
\beq                                                           \label{V-th}
             \Veff(\phi) = 2/z^2 + O(1),
\eeq
  and a contribution of the form $O(z^{-1})$ is absent. Thus we have an infinite potential wall that separates perturbations 
  to the left ($z< 0$) of the throat from those to the right ($z > 0$) because a natural boundary condition at $z=0$ is 
  $\psi(0) = 0$. All other perturbations also vanish at $z=0$. As a result, a mode of evident physical significance, connected 
  with time-dependent perturbations of the throat radius, drops off from the consideration.  

  A way to solve this problem is to apply a Darboux transformation (see, e.g., \cite{glamp} for a recent presentation and 
  application to BH physics), also called S-deformation, to the effective potential $\Veff$. It was used in \cite{kod03, kod11} 
  in order to convert a partly negative potential to a positive-definite one in a stability problem for higher-dimensional  \bhs. 
  Later on this method was used by Gonzalez et al. \cite{gonz1} for transforming a singular effective potential for perturbations
  of anti-Fisher \whs\ to a nonsingular one, which has led to finding an exponentially growing mode. Thus such \whs\ are
  unstable, the instability being connected with a changing radius of the throat. The same method in a slightly more general
  formulation was applied in \cite{sta1, sta2} in a stability study for other \sph\ configurations with throats, in particular, 
  some black universe models. The method can be briefly described as follows.

  Consider a wave equation of the type (\ref{wave})
\beq                                                        \label{wave1}
        \ddot \psi - \psi'' + W(z) \psi = 0,
\eeq
  with an arbitrary potential $W(z)$ (the above potential $\Veff$ is its specific example).  The potential $W(z)$ may 
  be presented in the form
\beq                                                        \label{W(S)}
        W(z) = S^2(z) + S',
\eeq
  which may be treated as a Riccati equation with respect to $S(z)$, and its solution makes it possible to find $S(z)$
  for given $W(z)$. Then we can rewrite \eqn{wave1} as 
\beq                                                        \label{wave2}
        \ddot\psi + (\d_z + S)(-\d_z + S)\psi =0.
\eeq
  If we introduce the new function
\beq                                                        \label{chi}
        \chi = (-\d_z + S)\psi,
\eeq
  then, applying the operator $-\d_z + S$ to the left-hand side of \eq (\ref{wave2}), we obtain a wave equation for $\chi$:
\beq                                                        \label{wave3}
        \ddot\chi - \chi'' + W_1 (z)\chi = 0,
\eeq
  with the new effective potential
\beq                                                        \label{W1}
        W_1 (x) = -S' + S^2 = - W(z) + 2 S^2.
\eeq

  If a static solution $\psi_s(z)$ of \eq (\ref{wave1}) is known, so that $\psi''_s = W(z)\psi_s$, then we can choose
\beq                                                       \label{S-sol}
        S(z) = \psi'_s/\psi_s
\eeq
  to use in the above transformation. Next, if we assume that $W(x) \approx az^{-n}$ at small $z$ and require that $W_1$ 
  should be finite at $z=0$, then, using \rf{W1}, we find that a necessary condition for removing such a singularity is 
  $n=1,\ a = 2$, that is, $W \approx 2/z^2$. Fortunately, according to \eqn{V-th}, the potential $\Veff$ exhibits precisely 
  the required behavior. One can notice that this regularization works for a positive pole, $W \to +\infty$ as $z\to 0$, and 
  removes a potential wall in $W(z)$ whereas a potential well cannot be removed in this way. A point of interest is that the 
  Darboux transformation, being isospectral when it connects regular wave equations, loses isospectrality in the singular case
  and actually reveals a mode of perturbations which was hidden when the wave equation had a singular potential.  

  From \rf{W1} with finite $W_1$ it also follows that at small $z$
\beq                                                             \label{S0}
    S \approx  - 1/z   \ \ \then \ \  \psi_s \propto 1/z.
\eeq
  
  Summarizing, we see that the singularity of the effective potential $\Veff$ occurring at a throat can be regularized for a throat 
  of generic shape \cite{gonz1, sta1}. What is also of great importance is that solutions to the regularized wave equation 
  lead to regular perturbations of the scalar field and the metric. It was this procedure that has led to a proof of the instability
  of anti-Fisher (Ellis type \cite{br73, ellis}) \whs\ \cite{gonz1} and other configurations with scalar fields in GR \cite{sta1, sta2}.
  Examples of such results are depicted in Fig.\,5:

\begin{figure*}
\centering
\includegraphics[width=13cm]{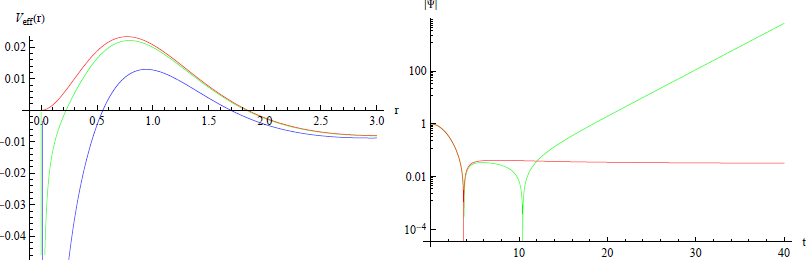}
\caption{\footnotesize
        Regularized effective potentials for perturbations of some black-universe solutions
        \rf{r1}, \rf{B1} (left) and the time-domain profiles of their time evolution \cite{sta2} (right). 
        Only one of the models under study is stable (the time-domain plot is asymptotically constant), 
	 it is the model where the throat coincides with the event horizon.} 
\end{figure*}

\subsection {Perturbations near the surface $h = 0$}

  The effective potential $\Veff$ also possesses another kind of singularities at values of the radial coordinate 
  where the factor $h(\phi)$ in (\ref{act}) changes its sign, if certainly such values of $\phi$ do exist, which happens in the 
  trapped ghost situation. The nature of these singularities is different from that described above. Somewhat similar singularities 
  were previously found  in systems with conformal continuations at the corresponding transition surfaces  
  \cite{br-JMP, stepan1, stepan2}. The latter phenomenon takes place, for example, in scalar-tensor theories of gravity where
  it can happen that at certain values of the parameters the whole  Einstein-frame manifold maps to only a region in the 
  Jordan-frame manifold (or vice versa). It was found that for \ssph\ configurations of this kind, monopole perturbations 
  obey wave equations of the form \rf{wave} with effective potentials possessing singularities of the form
\beq                                           
           \Veff  \approx - 1/(4z^2) \quad\ {\rm as} \quad z\to 0    \label{sing-}
\eeq   
  where $z=0$ is a transition surface on which the Einstein frame terminates while the Jordan frame has a regular metric 
  \cite{stepan1}.

  Returning to our system, let us assume that close to the value of $z$ at which $h=0$ (let it be $z=0$) the metric functions
  $\alpha = \gamma$ (according to our ``tortoise'' gauge) and $\beta$ as well as the scalar $\phi$ are regular and can 
  be approximated by the Taylor expansions
\bearr
              \alpha = \alpha_0 + \alpha_1 z + \ldots, \quad\ 
	     \beta = \beta_0 +  \beta_1 z + \ldots, \quad\ 
	    \phi = \phi_0 + \phi_1 z  + \half \phi_2 +\ldots,
\nnn
	     h(z) = h_1 z + \half h_2 z^2 + \ldots,  \qquad \phi_1\ne 0, \ \ \ h_1 \ne 0.
\ear
  One can show that in such a generic situation the potential \rf{Veff} is approximated near $z=0$ by the Laurent series 
\beq                                                                                         \label{Veff-h}
            \Veff (z) = - \frac{1}{4z^2} 
          + \frac{1}{z}\biggl[\frac{h_2}{h_1} + 2\Big(\beta_1 - \alpha_1
		+\frac{\phi_2}{\phi_1}\Big)\biggr]	+ O(1).      
\eeq

  Thus the surface where $h=0$ is a location of a potential well of infinite depth. In quantum mechanics such a well, according 
  to the stationary Schr\"odinger equation, would cause the existence of arbitrarily deep negative energy levels. As regards 
  the stability study, it would be tempting to conclude, by analogy, that there are perturbation modes with $\omega^2 < 0$ 
  and arbitrarily large $|\omega|$, and then the corresponding perturbations would grow as ($\df \sim e^{|\omega| t}$),
  immediately leading the system out of a linear regime and making necessary a nonlinear or nonperturbative analysis. 

  However, in the stability study, the quantities $\psi$ in \eqn{wave} or $Y$ in \rf{Schr} are subject to quite different physical
  requirements: the perturbation $\df$ should remain finite in the whole space and also vanish at infinities or horizons, while 
  in quantum mechanics the only requirement is quadratic integrability of $\psi$.
  
  With the potential \rf{Veff-h}, the general solution of \rf{Schr} at small $z$ has, independently of $\omega$, the leading terms
\beq
           Y(z) = \sqrt {|z|} (C_1 + C_2 \ln |z|) + O(z^{3/2}), \qquad C_1,\ C_2 = \const. 
\eeq   
  Furthermore, according to \rf{to_psi}, $\df \sim Y/\sqrt{h} \sim Y/\sqrt{|x|}$, and
\beq                                                \label{df-h}
              \df \sim C_1 + C_2 \ln |z| + O(z),
\eeq
  so physically meaningful perturbations correspond to $C_2=0$. The negative pole of  \rf{Veff-h} thus leads to a new 
  constraint that should be considered together with the ordinary boundary conditions specifying the boundary-value problem 
  for the \Schr\ equation. From this additional constraint it may follow that this boundary-value problem does not possess
  any discrete spectrum. Indeed, assuming a background configuration with a surface where $h=0$ (say, at $z=0$), 
  \eqn{Schr} has a solution satisfying the boundary conditions imposed at infinities and/or a horizon; then there will be in
  general a zero probability that such a solution is finite at $z=0$. A possible exception is a $\Z_2$-symmetric background 
  with respect to the surface $z=0$ (then the whole boundary-value problem is the same for $z<0$ and $z>0$, and the
  whole spectrum is also the same), but in our system it is manifestly not such a case since we have $h >0$ on one side of 
  the surface $z=0$ and $h <0$ on the other. A situation like this was discussed in \cite{b-kir,turok} for the stability of 
  BHs with a conformal scalar field \cite{we70, bek}, and after all, using some more reasoning, it was concluded that these 
  black holes are stable under monopole perturbations \cite{turok}.
  
  (Unlike this situation, in a quantum-mechanical problem setting, the quadratic integrability requirement for $Y(z)$ is satisfied 
  as long as $\int z^n \ln^2 z \,dz <\infty$ near $z=0$ for any $n \geq 0$. A quantum particle in such a potential well, 
  as is sometimes said, ``falls onto the center'' \cite{LLQ} since, as one descends to deeper and deeper negative energy 
  levels, the wave function becomes more and more concentrated near $z=0$.)    
    
  We can conclude that if there is a transition surface from canonical to phantom scalar field, it plays a strong stabilizing role
  for any such configuration. However, it looks impossible to say about a particular trapped-ghost solution, whether it is stable 
  or not: an individual investigation is necessary since a conclusion must depend on the details of each model.  

\begin{figure}
\centering
\includegraphics[width=6.5cm]{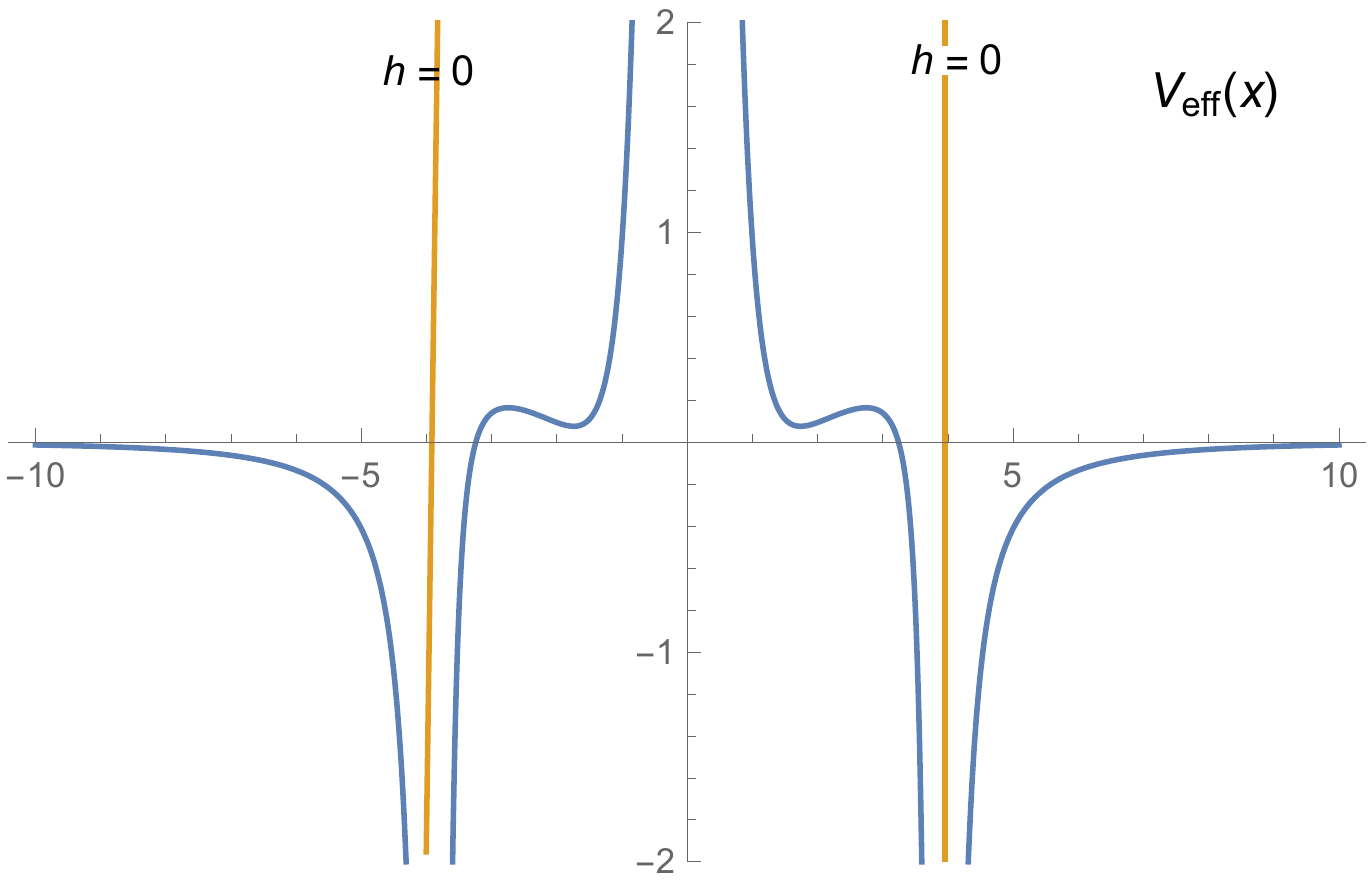}\qq
\includegraphics[width=6.5cm]{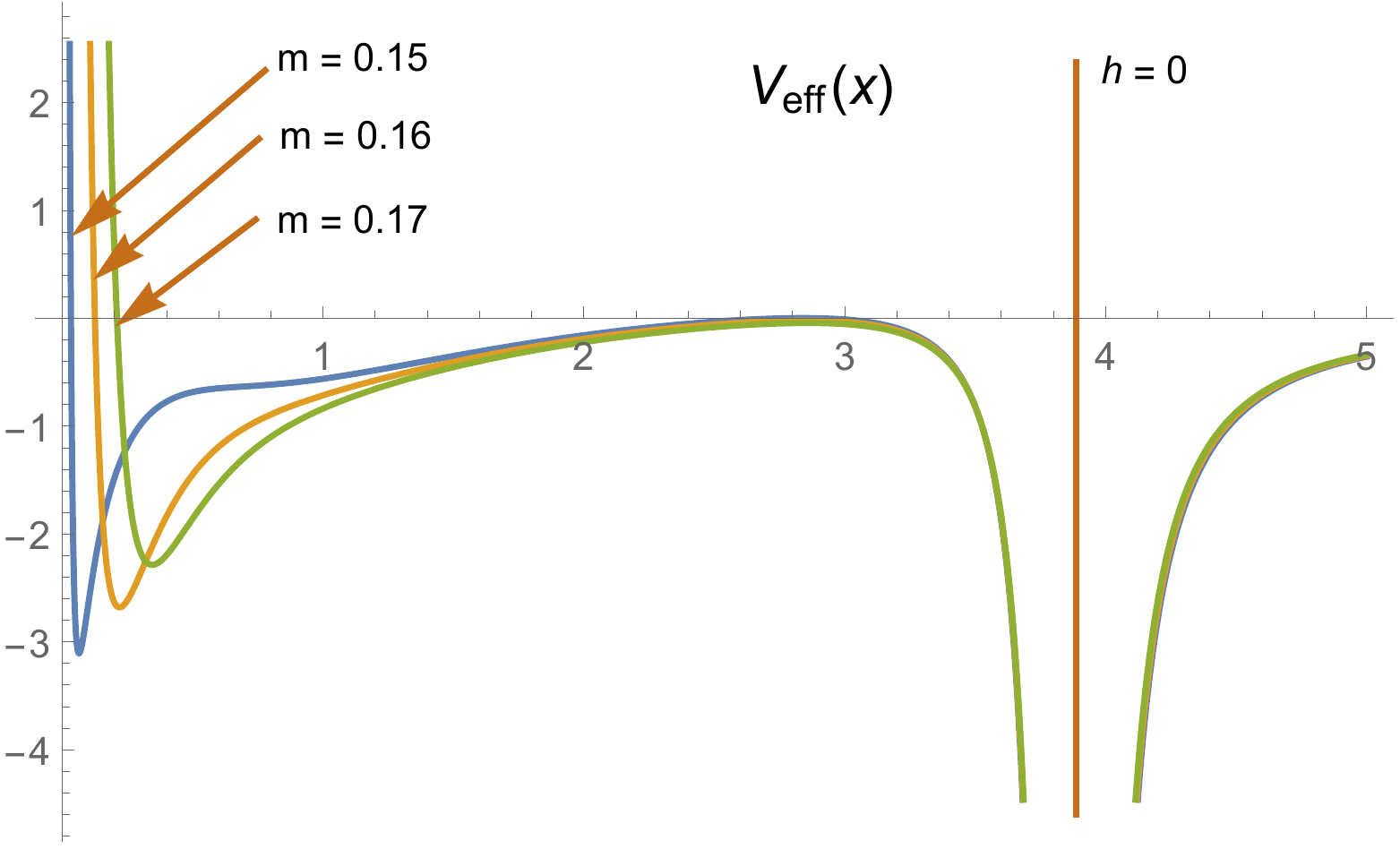}
\caption{\footnotesize
         Effective potentials $\Veff = \Veff (x)$ for perturbations of trapped-ghost solutions with the metric
	  coefficients \rf{r2} and \rf{B2}, $n=3$. Left: $\Veff$ for a massless ($m=0$) symmetric \wh.
	  Right: $\Veff$ for trapped-ghost black universe solutions with $m >0$, in which the minimum of 
	  $r$ ($x=0$) is beyond the horizon. Note that in terms of $z$, the first plot would remain qualitatively 
	  the same, whereas in the second one a finite $x >0$ at the horizon would turn into $z = -\infty$.}
\end{figure}

 Figure 6 shows the effective potentials $\Veff$ for different trapped-ghost configurations. In the case of a symmetric, twice 
 \asflat\ \wh\ there are four regions separated by poles of $\Veff$. The positive pole at $x=0$ can be regularized by an 
 S-transformation as described above, while the poles where $h=0$ impose conditions similar to $C_2=0$ in \eqn{df-h}.
 Thus, for \eqn{Schr}, as many as three boundary-value problems should be considered in three ranges of $z$ separated from 
 each other by $h=0$ points. Though, in this particular solution two of them coincide due to symmetry 
  ($x \leftrightarrow -x$) of the \wh.

 For black-universe solutions without throats outside the horizon, an additional regularization is unnecessary, two boundary-value
 problems on different sides of the pole at $h=0$ may be solved directly, and the resulting spectra should be then compared 
 to find out whether there are coinciding eigenvalues leading to unstable perturbation modes. A tentative conclusion is that 
 at least for some values of $m$ there is such an unstable mode with $\omega =0$, which means that there are perturbations
 growing with time by a linear law.    

\subsection{Perturbations near a long throat}

  Now, let us apply the above formalism to static solutions where the spherical radius $r(x)$
  behaves near the throat like the function \rf{r3}. More specifically, let the throat be located at 
  $z = 0$ and, as $z \to 0$, let the radius $r$ behave as 
  $r \approx r_0 + O(z^{2n})$ (as is the case with \rf{r3}).\footnote
	   {Since the coordinates $x$ and $z$ are related by $dz = dx/A(x)$, and $A(x)$ 
	     is finite at the throat, the function $r(z)$ qualitatively behaves in the same way as $r(x)$.}
  Then at small $z$, as can be directly verified \cite{invis2}, 
\beq                 \label{z0}
           \Veff (z) = 2\biggl[ \Big(\frac{\beta''}{\beta'}\Big)^2 - \frac{\beta'''}{\beta'}\biggr] + O(1)
                         = \frac{2(2n{-}1)}{z^2} + O(1),     
\eeq
  Models with ordinary (generic) throats correspond to $n=1$ and those with long throats to $n > 1$.

  As described above, for $n=1$ the potential $\Veff$ can be regularized using a Darboux transformation, which is a 
  special kind of substitution in \eqn{Schr}: the transformed equation is regular at $z=0$, and solutions to the corresponding 
  boundary-value problem describe regular perturbations of the scalar field and the metric, including those in which the 
  throat radius changes with time. However, a necessary condition for singularity removal is that $\Veff = 2/z^2 + O(1)$, 
  which does not hold for $n > 1$. 

  Now suppose that (without loss of generality) at $z=0$ the potential behaves as 
\beq
	W(z) = \frac{N}{z^2} + O(1), \cm  N = \const,
\eeq
  and apply the S-transformation according to \eqs \rf{wave1}--\rf{W1}. It is then easy to verify that 
\beq					\label{W1+}
          W_1(z) = \frac{N_1}{z^2} + O(1), \qq   N_1 = N+1 \pm \sqrt{1+4N}.  
\eeq
  The minus sign should be chosen here if we wish to ``weaken'' the pole of $W$, that is, make $N_1 < N$. In particular, 
  if $N = N_k = k(k+1)$, $k\in \N$, then the S-transformation leads to a new potential with $N_{k-1} = (k-1)k$, or symbolically
\beq
	N_k = k(k+1) \ \ \mathop{\longrightarrow}\limits_S \ \ N_{k-1} = (k-1)k.
\eeq
  Therefore, if a potential behaves as $W(z) = -k(k+1)/z^2 + O(1)$, $k \in \N$, then each described step 
  ``lowers the order'' $k$ by a unit, and after $k$ such steps we obtain a regular potential.

  For long throats with radii of the form $r \approx r_0 + \const\cdot z^{2n}$, the effective potential  \rf{z0} thus admits 
  regularization in $k$ steps if $2n = 2n_k = 1 + k(k+1)/2$. Thus, for $k = 1,2,3,4,5$, the corresponding values of 
  $n = n_k$ admitting this procedure are $n =1, 2,  7/2, 11/2, 8$, etc.  If the quantity $n$ in \eqn{z0} does not belong 
  to this sequence, the potential $\Veff$ cannot be regularized by S-transformations.

  Thus a general formalism for studying the stability of \wh\ models with long throats is ready, at least for $n$ belonging to 
  the sequence $n_k = 1/2 + k(k+1)/4$. However, a practical implementation of this formalism for particular models is difficult
  and requires significant numerical work (now in progress) since even if static background solutions are known analytically,
  the perturbation equations can be solved only numerically. 

  We have considered \cite{invis2}, as a tentative study, the stability properties of a configuration with a constant spherical
  radius $r(x)$, which may be called a maximally long throat. It is not a wormhole since there are no spatial asymptotics. 
  The corresponding static solution to the field equations reduces to the well-known Nariai metric \cite{nariai} with a constant 
  scalar field, and it has been explicitly shown that it is unstable under linear perturbations. Therefore, one can speculate that 
  a slowly varying radius near a throat does not stabilize a wormhole supported by a phantom scalar field.

\section {Conclusion}

  Having recalled some general properties of \ssph\ space-times in GR, in particular, those with a scalar field source, we have 
  discussed some examples of globally regular solutions, among which are wormholes and black universes. The latter 
  seem to be a viable possible description of a pre-inflationary epoch in cosmology and an attractive opportunity that
  the cosmic evolution began from a horizon instead of a singularity. 

  Some of the existing solutions confirm that scalar fields may change their nature from a canonical one to a ghost one in a 
  smooth way without leading to space-time singularities \cite{rubin04,trap1,trap2,trap3}. This circumstance widens the 
  possible choice of scalar field dynamics and, in addition, the possible formulations of scalar-tensor theories of gravity, 
  even though we have so far considered minimally coupled fields only.

  More specifically, the trapped ghost concept makes it possible to obtain \sph\ \wh\ and regular BH models with a ghost 
  behavior in a restricted strong-field region whereas outside it, where any observers can live, the scalar has usual properties.
  One can speculate that if such ghosts do exist in Nature, they are all confined to strong-field regions 
  (``all genies are sitting in bottles'').

  In addition to particular examples of exact solutions, some general properties of \ssph\ scalar field configurations
  have been revealed:
\begin{description}
\item[(i)]
    Trapped-ghost solutions to the field equations are only possible with nonzero potentials $V(\phi)$.
\item[(ii)]
    If the Einstein-scalar equations have a twice \asflat\ solution (be it of trapped-ghost nature or not), 
    then the Schwarzschild mass has different signs at the two infinities, while mirror ($\Z_2$)
    symmetry with respect to a certain surface is only possible if $m = 0$.
\item[(iii)]
    Transition surfaces between regions of canonical and phantom behavior of a scalar field create a potential well for \sph\
    perturbations, with a universal shape (see \eqn{Veff-h}) for all such models. The finiteness requirement for perturbations 
    on this surface forms an additional constraint that, being added to the standard boundary conditions, restricts the set of 
    possible solutions and thus plays a stabilizing role.
\end{description}

  This latter result may seem somewhat unexpected but, in my view, it can raise an interest in trapped-ghost models. 
  One can try to use this opportunity for solving various problems of gravitational physics and cosmology. 

  Of certain interest are the properties of perturbations near a long throat which is obtained in some of the solutions. More 
  generally, one can conclude that a general formalism for stability studies in the presence of throats and transition surfaces
  has been prepared, but studies of particular models lead to rather involved numerical tasks to be considered in the near future.
  Another task is to try to extend the present results to some alternative theories of gravity, such as scalar-tensor and $f(R)$ 
  theories. 
  
  One should also mention such a task of utmost importance in the present and future studies as the analysis of 
  observational properties of \whs, such as gravitational lensing and shadows
  (see \cite{lens1, lens2, shad} and references therein)
  as well as possible constraints on their existence that follow from any already known observational data 
  (\cite{Meng} and references therein).
 
\subsection*{Acknowledgments}

  I thank the colleagues from Yerevan University for wonderful hospitality during the conference.
  I am grateful to Milena Skvortsova, Sergei Bolokhov and Artyom Yurov for helpful discussions. 
  The work was partly performed within the framework of the Center FRPP 
  supported by MEPhI Academic Excellence Project (contract No. 02.a03.21.0005, 27.08.2013).
  This publication was also supported by the RUDN University program 5-100 and by RFBR grant 16-02-00602.


\end{document}